\documentclass[superscriptaddress,showkeys,twocolumn,nofootinbib,longbibliography]{revtex4-1}

\usepackage{amsmath}
\usepackage{amssymb}
\usepackage{graphicx}
\usepackage{epsf}
\usepackage{slashed}
\usepackage{enumitem}
\usepackage[english]{babel}
\usepackage[cp1250]{inputenc}
\usepackage{hyperref}
\usepackage{xcolor}
\usepackage{physics}

\usepackage[only,llbracket,rrbracket,llparenthesis,rrparenthesis]{stmaryrd} 
\usepackage{accsupp} 

\usepackage[font=small,labelfont=bf,justification=raggedright]{caption}
\usepackage{subcaption}

\usepackage{bbm} 

\usepackage[nodayofweek]{datetime}
\newdateformat{mydate}{\twodigit{\THEDAY}{ }\shortmonthname[\THEMONTH], \THEYEAR}
\usepackage[normalem]{ulem}

\usepackage{color}
\usepackage{ulem} 

\definecolor{myred}{rgb}{1,0,0}
\definecolor{mygreen}{rgb}{0,0.8,0.2}
\definecolor{myblue}{rgb}{0,0,1}

\definecolor{Ared}{rgb}{1,0.7,0}
\definecolor{Agreen}{rgb}{0.7,0.8,0.2}
\definecolor{Ablue}{rgb}{0,0.7,1}

\renewcommand{\emph}[1]{\textit{#1}}

\begin{document}


\title{From BPS geodesics to mode-driven dynamics in the\\ scattering of multiple BPS vortices}

\author{A. Alonso Izquierdo}

\affiliation{Departamento de Matem\'{a}tica Aplicada, University of Salamanca, Casas del Parque 2, 37008 Salamanca, Spain}
\affiliation{IUFFyM, University of Salamanca, Plaza de la Merced 1, 37008 Salamanca, Spain}

\author{M. Bachmaier}

\affiliation{Arnold Sommerfeld Center, Ludwig-Maximilians-Universit\"at, Theresienstra\ss{}e 37, 80333 M\"unchen, Germany}
\affiliation{Max-Planck-Institut fur Physik, Boltzmannstrasse 8, 85748 Garching, Germany}

\author{A. Wereszczynski}

\affiliation{Institute of Theoretical Physics, Jagiellonian University,
Lojasiewicza 11, Krak\'{o}w, Poland}
\affiliation{International Institute for Sustainability with Knotted Chiral Meta Matter (WPI-SKCM$^{\; 2}$), Hiroshima University, 1-3-1 Kagamiyama, Higashi-Hiroshima, Hiroshima 739-8526, JAPAN}

\begin{abstract}
We analyze how the geodesic motion in the 3- and 4-vortex sectors of the Abelian-Higgs model at critical coupling is deformed by the excitation of a massive bound mode. We find that the geodesics corresponding to BPS solutions with enhanced symmetry remain unchanged, although the direction of the actual motion depends on the mode-generated force, i.e., a force arising from the change of the mode frequency along the geodesic. In a generic case, for example in head-on collisions between the axially symmetric 1- and 2-vortex or between two 2-vortices, the vortex trajectories can differ strongly from the BPS geodesic. This enhances the chaotic behavior in the formation of the final state.
\end{abstract}

\maketitle

\section{Introduction}
For a long time, it has been believed that the dynamics of the vortices in the Abelian-Higgs model at the critical coupling,  that is, in the BPS limit, is fully captured by the geodesic motion on the moduli space of the energetically equivalent BPS solutions. In this moduli space approximation~\cite{Manton:1981mp} the vortices are subject only to the velocity-dependent force arising from a nontrivial curvature of the moduli space. This explains~\cite{Samols:1991ne} the well-known $90^\circ$ degree scattering in the head-on vortex-vortex collision~\cite{Ruback:1988ba, Shellard:1988zx, Myers:1991yh}. 

However, it has been recently shown that even for an arbitrarily small velocity, the geodesic motion can be strongly modified if an internal mode is excited. This has initially been studied in 2-vortex collisions~\cite{Krusch:2024vuy, AlonsoIzquierdo:2024nbn, Alonso-Izquierdo:2024fpw} and then continued in the 3-vortex sector, where the scattering of the unit charge vortices along the collinear and equilateral triangle geometries has been analyzed~\cite{Alonso-Izquierdo:2025suz}. In all cases, the excitation of a massive bound mode gives rise to a mode-generated force, whose strength and sign depend on the flow of the spectral structure, i.e. the change of the frequency of the mode on the moduli space. In particular, an attractive force emerging due to the excitation of the lowest mode leads to chaotic multi-bounce collisions. Excitation of a higher mode triggered a repulsive force which resulted in backscatterings. Importantly, the dynamics in the excited case was always {\it confined to the BPS geodesic} and no deviation from the geodesic path was observed in the mentioned publications.
This is because of the enhancement of the symmetry of the initial configuration, which simply excluded any other geometries of the 1-vortices.

In the present work, we analyze the scattering of excited BPS vortices in less symmetric configurations. For these cases, the precise BPS trajectory is not known and was therefore determined through numerical simulations. In particular, we will focus on the charge-3 sector, considering head-on collisions between a 1-vortex and an axially symmetric 2-vortex. We will excite the vortices and analyze how the trajectory changes under the influence of the mode-dependent forces. Afterwards, we will perform a similar analysis for several scenarios in the charge-4 sector.

In all cases without enhanced symmetry, we will clearly see that the excitation of a bound mode will have a significant impact on the evolution leading to a dramatic modification of the BPS geodesic.

\section{The model}

The Abelian-Higgs model (in 2+1 dimensions) is defined by the following Lagrangian density
\begin{equation}
\mathcal{L}=-\frac{1}{4} F_{\mu\nu}F^{\mu \nu} +
\frac{1}{2} \overline{D_\mu \phi}\, D^\mu \phi -\frac{\lambda}{8}
\left(\abs{\phi}^2-V^2\right)^2,
\label{action1}
\end{equation}
where $D_\mu \phi = \partial_\mu\phi -igA_\mu \phi$ is the covariant derivative and  $F_{\mu\nu}=\partial_\mu A_\nu - \partial_\nu A_\mu$ is the electromagnetic field strength tensor. Without loss of generality, we set the vacuum expectation value $V$ and the gauge coupling $g$ to one. Therefore, the model is fully parametrized by one dimensionless constant $\lambda$.
The corresponding field equations read as follows 
\begin{align}
    D_\mu D^\mu \phi + \frac{\lambda}{2}\phi \left(\abs{\phi}^2-1\right)&=0\, , \\
    \partial_\mu F^{\mu \nu} -\frac{i}{2}\left(\phi\, \overline{D^\nu\phi} - \overline{\phi}\, D^\nu \phi\right)&=0\, .
\end{align}
This theory has a charge-$n$ vortex solution, which is given by~\cite{Nielsen:1973cs} 
\begin{align}
\label{eq:vortex-ansatz}
  \phi^{(n)}(r,\theta)&= f_n(r) \, e^{in\theta}, \nonumber\\ A^{(n)}_{\theta}(r,\theta) &= n \frac{a_n(r)}{r}, \nonumber\\ A^{(n)}_r(r,\theta)&=0\, ,
\end{align}
where the profile functions $f_n(r)$, $a_n(r)$ obey the ordinary differential equations
\begin{align}
& \frac{\dd^2 f_n}{\dd r^2} + \frac{1}{r} \frac{\dd f_n}{\dd r} - \frac{n^2 (1-a_n)^2 f_n}{r^2} + \frac{\lambda}{2} f_n (1-f_n^2) = 0\, , \label{edo1} \\
& \frac{\dd^2 a_n}{\dd r^2} - \frac{1}{r} \frac{\dd a_n}{\dd r} + (1-a_n) f_n^2 = 0\, . \label{edo2}
\end{align}
The topologically non-trivial boundary conditions are 
\begin{align}
    f_n(r \to\infty) = 1\, ,&\hspace{1cm} \; a_n(r\to\infty)= 1\, ,\nonumber\\
    f_n(r\to0)=0\, ,&\hspace{1.19cm} \; a_n(r\to0)=0\, .
\end{align}
The charge-$n$ vortices are stable solitons for $\lambda \leq 1$. This includes the type-I regime ($\lambda < 1$), where unit-charge vortices attract each other and eventually merge into an axially symmetric, higher-charged vortex, as well as the BPS regime ($\lambda = 1$), in which the scalar and gauge forces between vortices exactly cancel.

\section{Geodesic motion}

In the BPS limit, due to the absence of static forces, a configuration with multiple unit-charge vortices placed at arbitrary positions is static, leading to a large family of energetically equivalent BPS solutions. They obey the following Bogomolny equations~\cite{Bogomolny:1975de}
\begin{align}
D_1\phi \pm i D_2 \phi &=0 \, ,\\ \hspace{0.5cm} F_{12}\pm
\frac{1}{2} \left(\abs{\phi}^2 -1\right)&=0\, , \label{fopdes}
\end{align}
and are saturating the topological bound on the energy, $E[\phi,A]= \pi |n|$.

This family is characterized by $n$ continuous parameters $z_k=x_k+iy_k$ with $k=1,\ldots, n$, which represent the positions of the zeros of the scalar field identified with the positions of the constituent unit-charge vortices~\cite{Weinberg:1979er, Taubes:1979tm}. These parameters are called moduli and can be treated as coordinates on the moduli space (space of energetically equivalent solutions). This curved space is equipped with a metric inherited from the usual $L^2$-metric in the infinite-dimensional Euclidean space of field configurations. In practice, the metric $g_{ij}(z_k)$ with $i,j=1,\ldots,2n$, is found by inserting the BPS solution into the Lagrangian and performing the integration over the spatial variables. One should also impose the Gauss law which guaranties that the time variation of the fields is orthogonal to the gauge orbit.

Let us begin with the charge $n=3$ sector. The corresponding moduli space $\mathcal{M}_3$ has a rather non-trivial curved geometry. Since the information of the moduli space comes from the positions of the unordered zeros of the scalar field, it can be equivalently encoded in the roots of a third-degree complex polynomial
\begin{align}
        P_3(z)&=(z-z_1)(z-z_2)(z-z_3)\nonumber \\
        &= z^3+w_2z^2+w_1z+w_0\, ,
\end{align}
where
\begin{align}
    w_2&=-(z_1+z_2+z_3)\, , \\
    w_1&=z_1z_2+z_2z_3+z_3z_1\, , \\
    w_0&=-z_1z_2z_3\, .
\end{align}
Since the position of the center of mass decouples from the other degrees of freedom, we can fix it to be at the origin, $w_2=0$. This leads to the two-dimensional moduli space $\mathcal{M}_3^{\rm{CM}}$ defined by the polynomials
\begin{equation}
    P_3^{\rm{CM}}(z)= z^3+w_1z+w_0\, .
\end{equation}

\begin{figure}
\centering
    \includegraphics[width=1.0\linewidth]{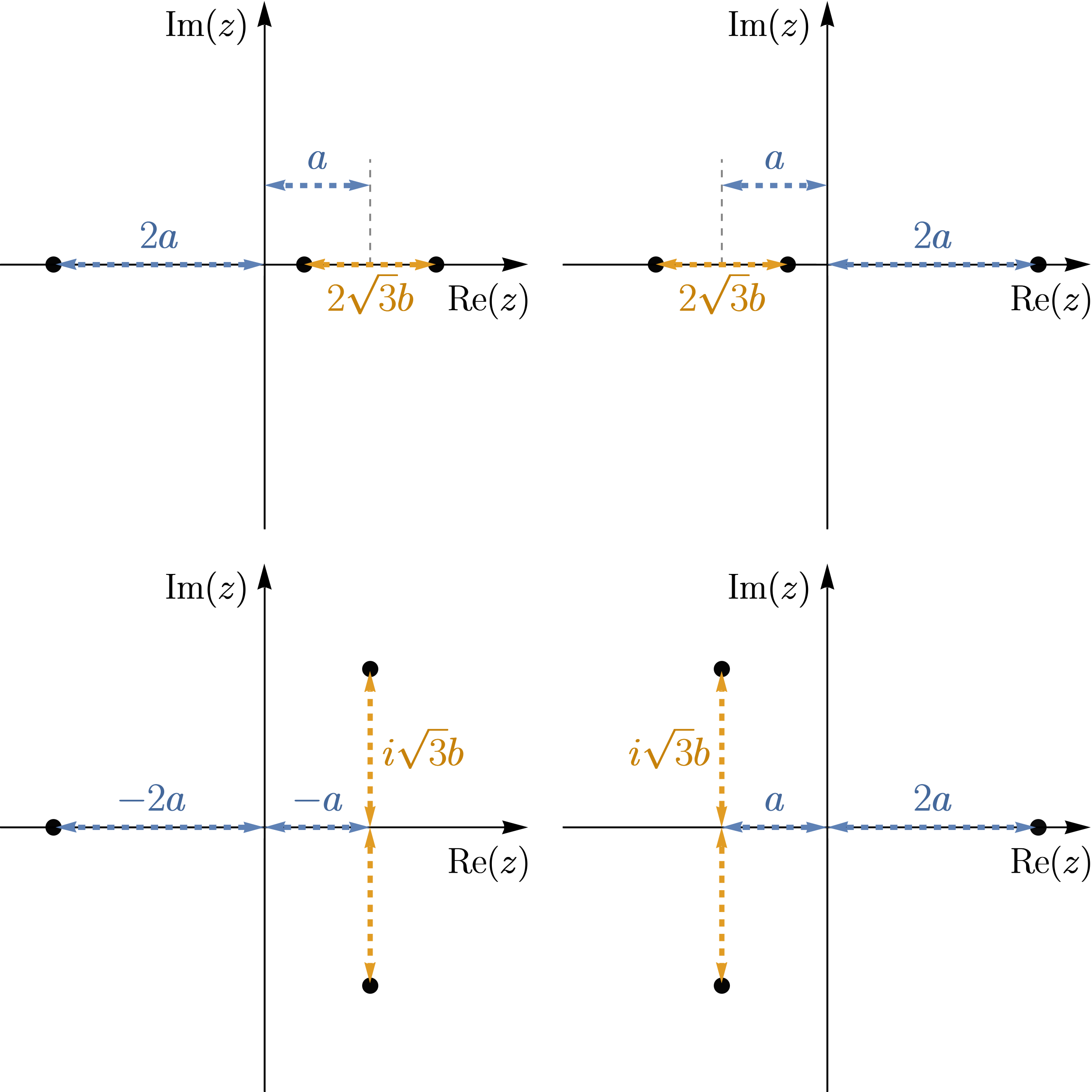}
    \caption{This figure illustrates the definition for the values $a$ and $b$ for the four different quadrants in the $a$-$b$-plane.} 
    \label{fig:explanation-parametrization-1+2}
\end{figure}

By assuming that $w_1$ and $w_0$ are real, this space can be further reduced to a two-dimensional subspace.
This is equivalent to considering only those vortex solutions that are invariant under the reflection symmetry $y \to -y$. 
The resulting space is a complete, geodesic submanifold $\widetilde{\mathcal{M}}_3^{\rm CM}$.
A useful choice of coordinates for the position of the zeros is given by~\cite{Alonso-Izquierdo:2025suz}
\begin{equation}
    z_1=2a\, , \;\; z_2=-a+\sqrt{3}b\, , \;\; z_3=-a-\sqrt{3}b\, , \label{zeros}
\end{equation}
with $a\in \mathbb{R}$ and $b \in \mathbb{R}_+ \cup i \mathbb{R}_-\cup \{0\}$. The values $b \in \mathbb{R}_+$ correspond to vortex configurations in which all vortices are collinearly arranged along the $x$-axis (see Figure~\ref{fig:explanation-parametrization-1+2} (upper plots)). Two vortices located off this axis are described by the values $b\in i \mathbb{R}_-$ (see Figure~\ref{fig:explanation-parametrization-1+2} (lower plots)). 
In this parametrization $w_1$ and $w_0$ are real numbers and read
\begin{equation}
   w_1=-3(a^2+b^2)\, , \;\;\; w_0=-2a (a^2-3b^2)\, .
\end{equation}
To avoid double counting, it is necessary to remove the region $b> \abs{a}/3$, which is a cone in the $a$-$b$-plane with an opening angle of $2\pi/3$. 
The boundaries of the excluded region, $b=a/\sqrt{3}$ for positive $a$ and $b=-a/\sqrt{3}$ for negative $a$, lead to exactly the same solutions
\begin{align}
    & z_1=2a\, , \;\;z_2=0\, , \;\;z_3=-2a\, , \;\; \mbox{for} \;\; a>0\, ,\nonumber \\
    & z_1=2a\, , \;\; z_2=-2a\, , \;\; z_3=0\, , \;\; \mbox{for} \;\;a<0\, .
\end{align}
Notice that we can always relabel the vortices, which leads to the same solution.

\begin{figure}
\centering
    \includegraphics[width=1.0\linewidth]{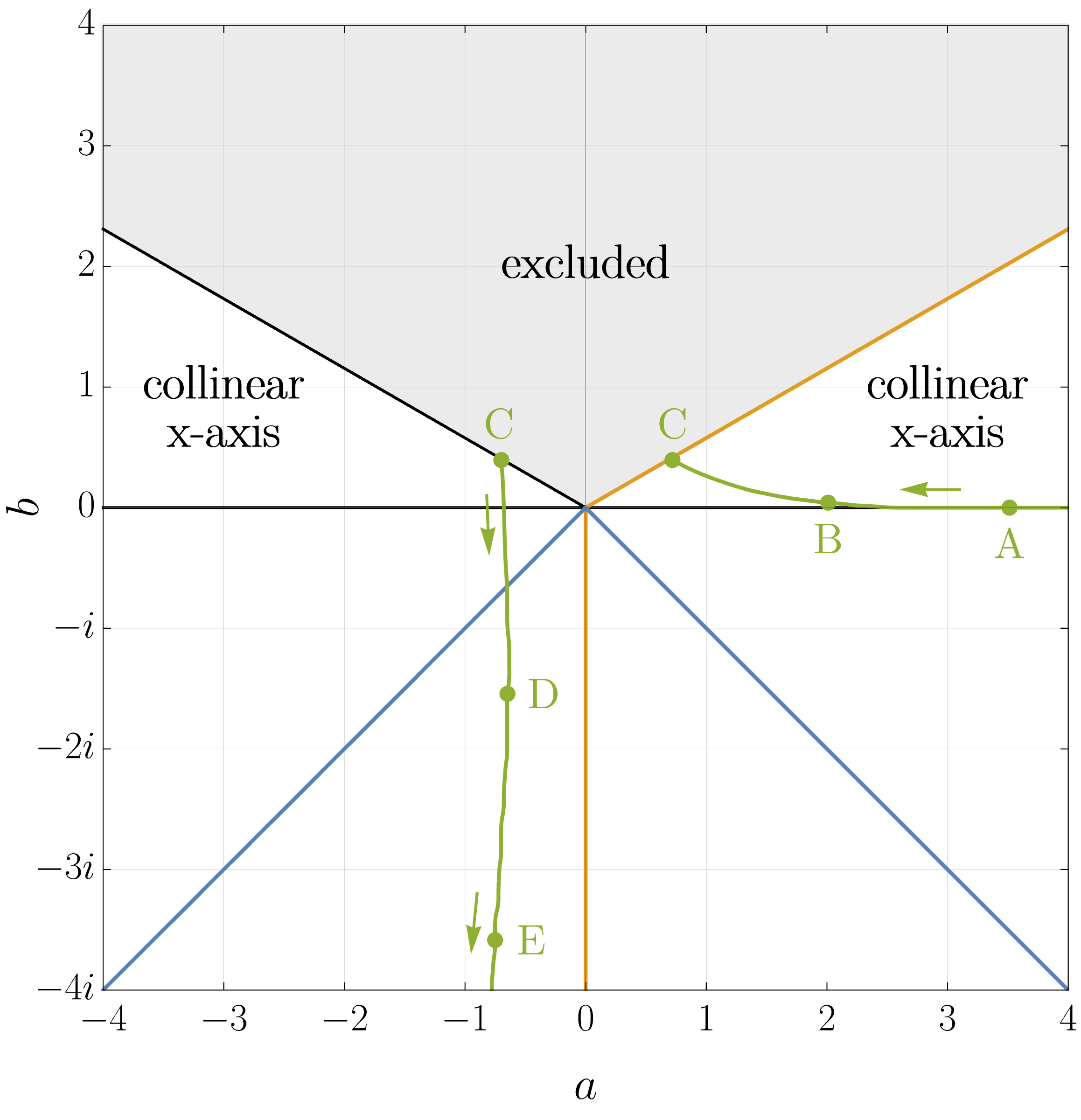}
    \caption{This plot shows the moduli space with the geodesics corresponding to the equilateral triangle (blue), collinear (orange), and 2+1 collisions (green). The labeled dots correspond to the time frames in Figure~\ref{fig:energy_density_1+2}.
    The 2+1 collision was obtained from a numerical simulation with initial vortex velocities $v_2 = 0.01$ for the 2-vortex and $v_1 = -0.02$ for the 1-vortex, respectively.}
    \label{fig:moduli_space}
\end{figure}

There are two natural submanifolds of one real dimension. Both possess an enhanced symmetry~\cite{Alonso-Izquierdo:2025suz}. The first, $\mathcal{N}^1$, is symmetric under the $y\to -y$ and $x\to -x$ reflections and corresponds to collinear configurations. This submanifold can be parametrized by
\begin{align}
    b&=\frac{a}{\sqrt{3}}\,, \;\; \mbox{for}\;\; a\in \mathbb{R}_+\, ,\nonumber \\
    a&=0\, , \;\; \mbox{for}\;\; b\in i\mathbb{R}_-\, ,\;
\end{align}
and encodes a head-on collinear collision of three 1-vortices. For $a>0$, the two outer vortices approach along the $x$-axis the inner vortex is located at the origin. They form the axially symmetric on-top configuration for $a=0$ and subsequently split along the $y$-axis for purely imaginary $b$-values. Thus, as in the case of the head-on collision of two 1-vortices we find a $90^\circ$ scattering. This geodesic is plotted in Figure~\ref{fig:moduli_space} (orange curve).

The second class of configurations with enhanced symmetry is described by the submanifold $\mathcal{N}^2$, which is invariant under the reflection $y \to -y$ and under $\mathcal{C}_3$ rotations, that is, rotations by $120^\circ$.
Here, the vortices are located at the vertices of an equilateral triangle 
\begin{equation}
    z_1=2a\, , \;\; z_2=-a+i\sqrt{3}a\, , \;\; z_3=-a-i\sqrt{3}a\, ,
\end{equation}
where $a \in \mathbb{R}$. This means that $b=ia$. The triangle rotates by $60^\circ$ as $a$ flips the sign ~\cite{Arthur:1995eh}. The corresponding geodesic is depicted in Figure~\ref{fig:moduli_space} (blue curve).

There is another, physically very well motivated scattering which does not correspond to an obvious geodesic path. Namely, a head-on collision between a 1-vortex and a 2-vortex. Such a two-body collision is much more natural than a three-body one. Furthermore, it is relevant in the near-BPS type-I regime, where an axially symmetric 2-vortex is a stable solution. It should be stressed that this case is not related to any enhancement of the symmetry.
Therefore, we have to use numerical methods to determine the geodesic path. More details about the numerical simulation are given in the Appendix.

As an initial state, we place the 2-vortex at $x=-7$ and the 1-vortex at $x=14$. The vortices are boosted toward each other with the small velocities $v_{2}=0.01$ and $v_1=-0.02$. Notice that the center of mass is at the origin and does not move during the scattering process. During the time evolution, we track the location of the vortices, i.e. the zeros of the scalar field, which are parameterized by~\eqref{zeros}, where now $a(t)$ and $b(t)$ are functions of time. This allows us to find the curve  $b(a)$, which due to the slow velocities can be treated as a geodesic path in moduli space.

Specifically, the zero of the charge-1 vortex coming from the right is at $z_1=2a$, while the two zeros of the 2-vortex coming from the left are at
\begin{equation}
    z_2=-a+\sqrt{3}b\,, \;\; z_3=-a-\sqrt{3}b\, .
\end{equation}  
Initially, $b=0$ and $a \to +\infty $ \footnote{In the numerical simulations by necessity, in the initial state we have the vortices separated by a finite (but large) distance. Due to the exponentially suppressed inter-vortex force and finite size of the grid, the boosted 2-vortex does not split for a small amount of time, which is consistent with the boost of the 2-vortex without any 1-vortex companion.}.

\begin{figure*}
    \centering
    \includegraphics[width=\textwidth]{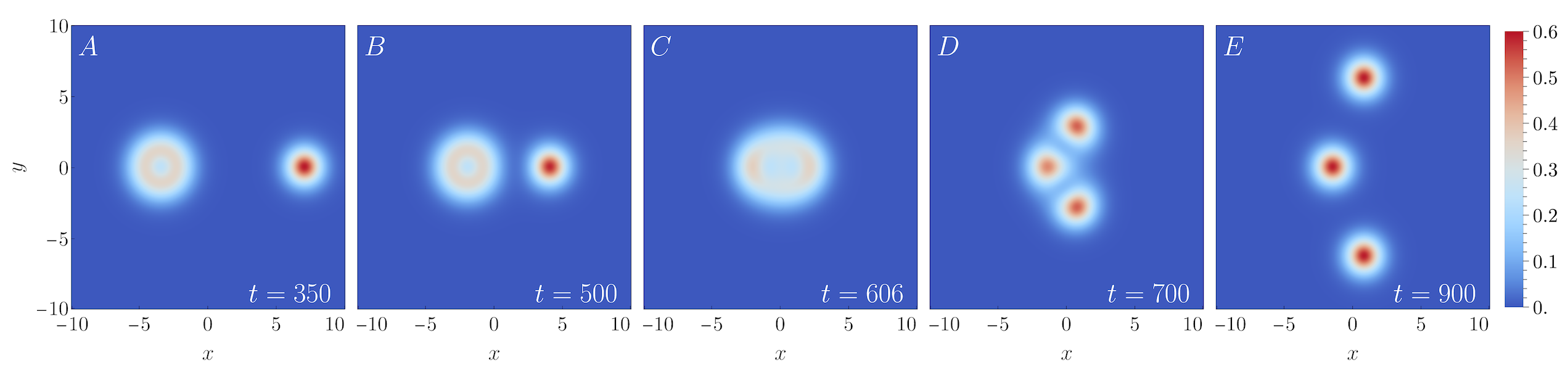}
    \caption{This figure displays five time frames of the energy density during the scattering of a $2$-vortex with a $1$-vortex. The animated version of the time evolution can be found in the following video: \url{https://youtu.be/o-0fQn-0wlE}}
    \label{fig:energy_density_1+2}
\end{figure*}

The final result of the geodesic is illustrated by the green line in Figure~\ref{fig:moduli_space}. Some time frames of the energy density of this simulation are given in Figure~\ref{fig:energy_density_1+2}.

In our numerical simulation, we found that at the beginning the 2-vortex splits along the $x$-axis while approaching the 1-vortex. At this stage of the evolution, all three zeros lie on the $x$-axis. This means that as the positive $a$ decreases, $b$ grows. The corresponding geodesic rises to the upper plane of the schematic moduli space (see the green line in Figure~\ref{fig:moduli_space}). Then, for $a\approx 0.7068$ and $b\approx 0.4080$, it hits the boundary of the excluded region, where $b=a/\sqrt{3}$. Here, the vortices are located at 
\begin{equation}
    z_1=2a\,, \;\;z_2=0\,, \;\;z_3=-2a\, .
\end{equation}
This parameterization continues on the other side of the excluded region, which is equivalent to relabeling the vortices, $z_1\leftrightarrow z_3$, where now $a$ is negative. The zeros remain still on the $x$-axis, with the two right vortices approaching each other. This means that $b$ decreases. Finally, $b$ crosses zero and passes to imaginary values, which corresponds to the head-on collision of the right two vortices and, consequently, to their $90^\circ$ scattering. The parameter $a$ is now negative and for a long time it changes only slightly. Eventually, it slowly decreases to $a\to-\infty$. Thus, we observe that the most left vortex (located on the $x$-axis) moves to the left with a very small velocity.

The overall effect of the 2+1 scattering is a break up of the 2-vortex by the incoming 1-vortex. One of the emerging 1-vortices continues to go along the $x$-axis, while the other two are ejected off this axis. This happens in two stages: first, the breakup of the $2$-vortex into two $1$-vortices, and subsequently, a head-on scattering of one of them with the incoming 1-vortex.

\section{Spectral Flow along the geodesic}
\label{sec:spectral-flow-along-the-geodesic}
As already studied in~\cite{Krusch:2024vuy, AlonsoIzquierdo:2024nbn, Alonso-Izquierdo:2024fpw} for two vortices and in~\cite{Alonso-Izquierdo:2025suz} for equilateral and collinear 3-vortex collisions, initial excitations of the vortices can change the dynamics by a lot. Therefore, to fully understand the 2+1 vortex collisions, it is necessary to study the flow of the bound modes on the geodesic.
Specifically, we consider the geodesic $b(a)$ that encodes the 2+1 collision and numerically find the static 3-vortex solutions corresponding to these particular locations of the zeros of the scalar field. Then, we consider small fluctuations around such a solution. This leads to an eigenvalue problem, which needs to be solved numerically. For details, we refer to \cite{Alonso-Izquierdo:2023cua, Alonso-Izquierdo:2024bzy}.

In the initial state, the separation between the 1- and 2-vortex is large. At such distances, the structure of the modes is well understood, since the 1-vortex and 2-vortex can be treated independently. The unit charge vortex has only one massive bound mode, with frequency $\omega_{10}^2=0.77747$ and angular momentum $k=0$. The axially symmetric 2-vortex has a more involved spectral structure~\cite{Goodband:1995rt, Alonso-Izquierdo:2024bzy}. The lowest mode is a massive bound mode with zero momentum and frequency $\omega_{20}^2=0.53859$. In addition, there are two degenerate modes with angular momentum $k=1$ and frequency $\omega^2_{21}=0.97303$. 

In the final state, there are three infinitely separated 1-vortices. Each of them has one bound mode $\omega_{10}$. The normalized orthogonal eigenstates generated by the normalized shape modes of the three 1-vortices, $\ket{i}$, are given by~\cite{Alonso-Izquierdo:2025suz}
\begin{align}
    \xi_1=&\frac{1}{\sqrt{3}} \left(\ket{1}+\ket{2}+\ket{3}\right)\, , \\
    \xi_2=&\frac{1}{\sqrt{2}} \left(\ket{1}-\ket{3} \right)\, , \\
    \xi_3=&\frac{1}{\sqrt{6}} \left(\ket{1}-2\ket{2}+\ket{3} \right)\, .
\end{align}
Here, the minus signs in front of the mode states correspond to a phase shift $i\omega t \mapsto i\omega t + i\pi$ relative to the mode states with the plus sign.

As we travel along the 2+1 geodesic, we should see a transition between the initial- and the final-state modes. Since the number of modes changes it means that at least one mode that exists in the initial state has to disappear, or, speaking precisely, pass to the continuum spectrum, during the collision.

\begin{figure}
    \centering
    \includegraphics[width=1.0\linewidth]{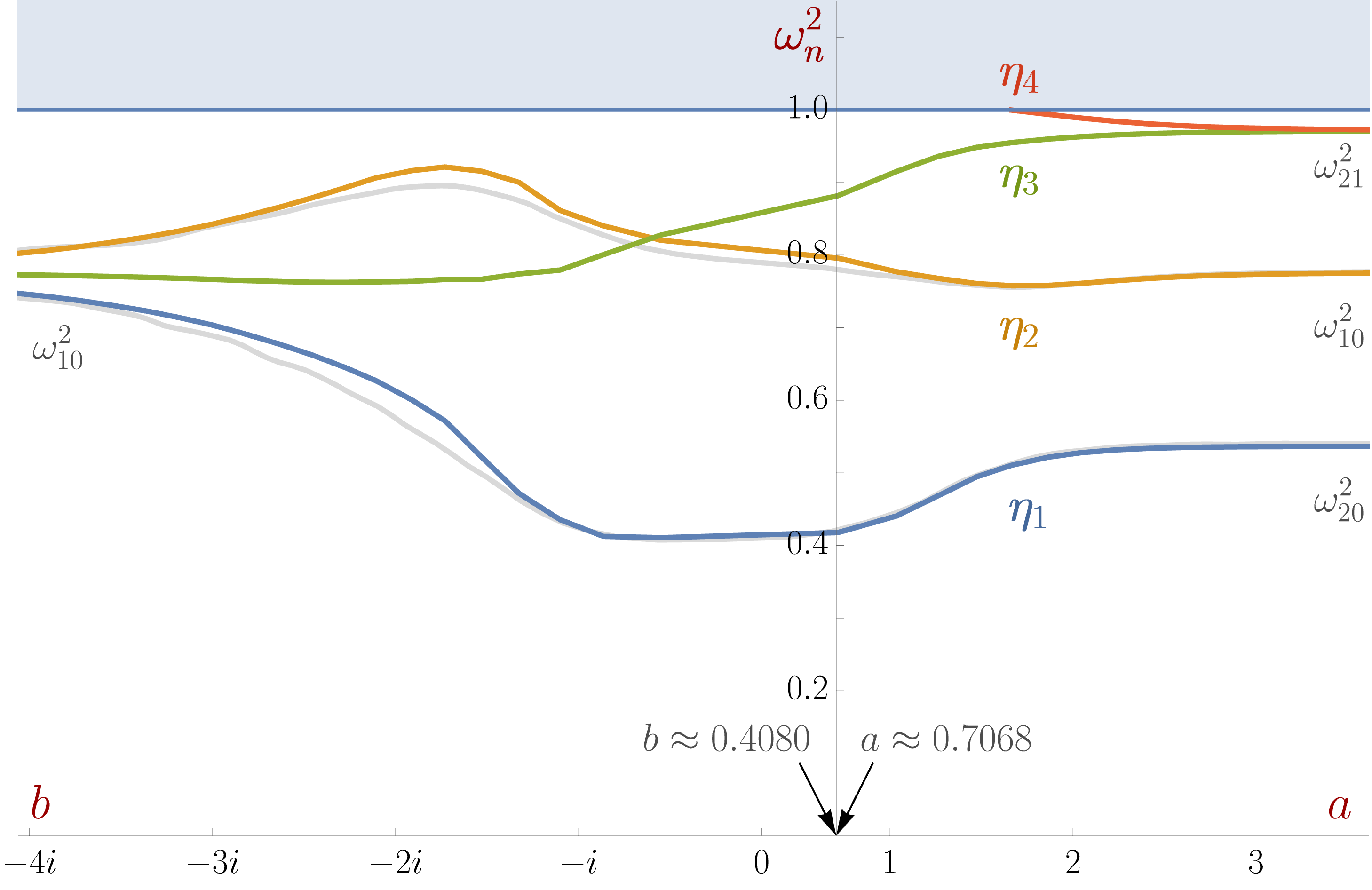}
    \caption{This figure shows the flow of the spectrum along the BPS geodesic path of the 2+1 collision. The r.h.s. of the frequency axis corresponds to the part of the geodesic with $a>0$. The l.h.s. of the frequency axis corresponds to the part of the geodesic with $a<0$. The light-gray lines show the measured spectral flow obtained from the numerical simulation.}
    \label{fig:flow}
\end{figure}

The actual flow of the modes along the geodesic path is plotted in Figure~\ref{fig:flow}. Here, the right hand side of the horizontal axis covers the part of the geodesic with $a\in(0.7068,\infty)$, that is, before it hits the excluded region. The left hand side covers the part of the geodesic located in the region with $a<0$. Here, the geodesic can be uniquely parametrized by the coordinate $b$. It starts at $b=0.4080$ and then decreases to $0$. Subsequently, $b$ passes to negative imaginary values as the vortices begin to separate along the $y$-axis.

Together, there are four modes for the 2+1 configuration. The lowest mode $\eta_1$ (blue curve in Figure~\ref{fig:flow}) smoothly connects the lowest mode of the 2-vortex with the $\xi_1$ mode. The mode $\eta_2$ (yellow curve) interpolates between the mode of the 1-vortex and the $\xi_3$ mode. The last two modes $\eta_{3,4}$ initially are degenerate. The $\eta_3$ mode (green curve) transits in the $\xi_2$ mode, while the $\eta_4$ mode (red curve) hits the mass threshold. This occurs for a collinear geometry for $a\approx 1.6651$ and $b\approx 0.0867$. As is known, such a transition of the mode to the continuous spectrum triggers the spectral wall phenomenon~\cite{Adam:2019xuc, Alonso-Izquierdo:2024fpw}.

In addition, there is a level crossing at $b \approx -0.5489 i$. This is due to the fact that the 2+1 geodesic crosses the geodesic of the equilateral triangle scattering exactly for this value of $b$ (see the crossing of the yellow and green lines in Figure~\ref{fig:moduli_space}). 
Since two modes are always degenerate in the equilateral geometry~\cite{Alonso-Izquierdo:2025suz}, the flow of the modes along the 2+1 geodesic must contain a level crossing at this point.

\section{Deformation of the BPS geodesic motion}

The main insight derived from the spectral flow is the appearance of a {\it mode-generated force}. This is because excitation of the internal mode amounts to the appearance of a potential energy, which at the adiabatic approximation reads \cite{AlonsoIzquierdo:2024nbn}
\begin{equation}
    E_{\rm mod} = \frac{1}{2}\, C^2\,\omega (\vec{X})\, ,
\end{equation}
where $C$ is an adiabatic invariant and $\omega$ is the mode frequency that depends on the position on the moduli space, $\vec{X} \in \mathcal{M}$. Here $\vec{X}=(a,b) \in \widetilde{\mathcal{M}}_3^{\rm{CM}}$. 
Due to the position-dependence, this may result in a deformation of the trajectory on the moduli space.

Furthermore, there is the {\it Coriolis effect} as a second source for the deformation of the BPS geodesic, which describes a modification of the geometric, velocity-dependent force by the coupling between the zero and massive modes~\cite{Rawlinson:2019xsn}. 
On the level of a collective coordinate model, which contains both the zero mode (corresponding to the positions of the zeros of the vortices) and massive modes (corresponding to the amplitude of the shape modes), the Coriolis effect means a deformation of the moduli metric by amplitude-dependent terms
\begin{equation}
    g_{ij}(\vec{X}) \rightarrow g_{ij}(\vec{X}, \xi)\, ,
\end{equation}
where $\xi$ is the amplitude of the mode\footnote{Of course, since we included the amplitude of the mode as a new collective coordinate, the moduli space has one dimension more. Thus, the metric also has a component in the direction of the mode amplitude $g_{\xi \xi}$.}.
This has recently been computed in the 1-vortex case~\cite{Miguelez-Caballero:2025xfq} and results in a change of the constant metric function $g_{11}=\pi$ (mass of the 1-vortex) to a second order polynomial in the amplitude of the shape mode $g_{11}=\pi+c_1\xi+c_2\xi^2$, where $c_{1,2}$ are constants. We remark that this modification of the metric is essential for the correct description of some strongly non-BPS processes, e.g., kink-antikink collisions in 1+1 dimensions~\cite{Manton:2021ipk}.

In a generic situation, both the mode-generated force and the Coriolis effect influence the evolution, leading to a strong deformation of the BPS dynamics. However, in some special circumstances, there is an enhancement of the symmetry in the initial state, which has to be respected during the time evolution. This occurs in the head-on collision of two vortices as well as in the collinear and equilateral triangle collision of three vortices. For these cases, the scatterings, even in the excited state, must follow the BPS geodesic. This we will call  {\it quasi-BPS geodesic motion}. This means that the dynamics still follows the BPS geodesic path, but the actual time-dependent position on the BPS geodesic differs from the pure geodesic motion. The mode-generated force can slow down or speed up the motion along the geodesic. It can even change the direction of the motion, leading to backscattering or multi-bounce solutions. The quasi-BPS geodesic motion for the mentioned cases was discussed in detail in~\cite{Krusch:2024vuy, AlonsoIzquierdo:2024nbn, Alonso-Izquierdo:2024fpw, Alonso-Izquierdo:2025suz}.

It is important to underline that the BPS geodesic of the 2+1 scattering does not have any additional symmetry that forces the vortices to remain on the BPS geodesic. Therefore, the mode-generated force and the Coriolis effect will always deform the BPS geodesic and the trajectory of the vortices is allowed to explore any region of the $a$-$b$-plane for any non-zero value of the amplitude of the mode. This factor significantly increases the complexity of the dynamics.
Nevertheless, it is plausible to assume that the small amplitudes will only weakly modify the BPS geodesic. The flow of the mode along the deformed geodesic will differ only slightly from the flow found along the BPS geodesic. Due to that, we can still draw some conclusions about the modification of the dynamics. One should be aware that the transition from the quasi-BPS to non-BPS dynamics has a rather gradual and smooth character. 

The lowest mode, $\eta_1$, contributes to an attractive interaction (see the blue curve in Figure~\ref{fig:flow}). 
Here, initially $\eta_1$ is just the lowest mode of the axial 2-vortex. Then, it smoothly transits in one of the superpositions of the shape modes of three 1-vortices. That is, into $\xi_1$. The frequency of the $\eta_1$ mode decreases as $a$ decreases, that is, as the 2- and 1-vortex approach each other. The flow forms a (plateau-like) minimum around the point where the collinear configuration goes into a solution with two vortices located off the $x$-axis. Therefore, this configuration is energetically preferred and acts as an attraction point once the mode is excited. This suggests that for a sufficient high amplitude of the $\eta_1$-excitation, multi-bounce solutions may exist.

This bounce structure is triggered by the {\it resonant energy transfer mechanism}~\cite{Sugiyama:1979mi, Campbell:1983xu, Manton:2021ipk} (see also~\cite{Bachmaier:2025igf} for its role in vortex-antivortex collisions). In every collision, the energy flows between the kinetic and the internal degrees of freedom. Sometimes, the kinetic energy can be large enough to allow to overcome the mode generated attraction. This results in the escape of the vortices to infinity.

An excitation of the $\eta_2$ mode along the BPS geodesic (orange curve in Figure~\ref{fig:flow}), leads to a repulsive force. Initially, this mode is the shape mode of the 1-vortex and is smoothly connected with the $\xi_3$ mode of the three 1-vortices. The frequency increases, which results in repulsion. Thus, if this mode is excited, the actual dynamics can lead to backscattering. Multi-bounce solutions are not possible.

Similarly, excitation of the $\eta_3$ mode, the green curve in Figure~\ref{fig:flow}, leads to an attractive force. This mode interpolates between the higher mode of the axial 2-vortex and the $\xi_2$ mode. In this case, the frequency is a monotonously decreasing function as we go along the geodesic. Therefore, as long as the BPS geodesic path is not too much deformed, in the excited version, there are no backscattering or multi-bounces.

\section{Collisions of excited vortices}

In this section, we present our numerical tests for the above prediction. Concretely, we scatter the 2-vortex with the 1-vortex with a specific mode excited. Additionally to the figures presented in this section, you can find the animated results in the following video:\\
\url{https://youtu.be/o-0fQn-0wlE}

For a single charge-$n$ vortex, the lowest mode excitations can be added by
\begin{align}
    \phi^{(n)}(r,\theta,t)&=f_n(r)\, e^{in\theta}+\xi\, u_n(r)\, e^{i\omega_n t}\, e^{in\theta}\, ,\\
    A_\theta^{(n)}(r,\theta,t)&=n\frac{a_n(r)}{r}-\xi\, v_n(r)\, e^{i\omega_n t}\, ,
\end{align}
where $\xi$ is the initial amplitude of the mode. 
The linearized equations (in the background gauge~\cite{Cheng:1984vwu}) for the mode profile functions $u(r)$ and $v(r)$ are given by
\begin{align}
 &\omega^2 u_n= -\frac{\dd^2u_n}{\dd r^2} - \frac{1}{r} \frac{\dd u_n}{\dd r}\, + \nonumber\\  
 &+ \left(\frac{n^2(1-a_n)^2}{r^2} +\frac{3\lambda}{2} f_n^2 -\frac{\lambda}{2}\right)u_n+\frac{2n}{r} (1-a_n)f_n v_n\, ,
 \label{fluctuation-2}\\
  &  \omega^2 v_n=-\frac{\dd^2v_n}{\dd r^2} - \frac{1}{r} \frac{\dd v_n}{\dd r}\,+\nonumber \\ &+\left(\frac{1}{r^2} +f_n^2\right) v_n +\frac{2n}{r} (1-a_n)f_n u_n\, . 
  \label{fluctuation-1}
\end{align}
The mode function were computed numerically and normalized by
\begin{align}
    \int \dd^2 x \left(v_n(r)^2+u_n(r)^2\right)=1\, .
\end{align}
In the following we will consider the 1-vortex mode with a frequency of $\omega_{10}^2=0.77747$ and the lowest mode of the 2-vortex with a frequency of $\omega_{20}^2=0.53859$.

\subsection{$\eta_1$ mode}

\begin{figure}
    \centering
   \hspace*{-0.5cm} \includegraphics[width=1.03\linewidth]{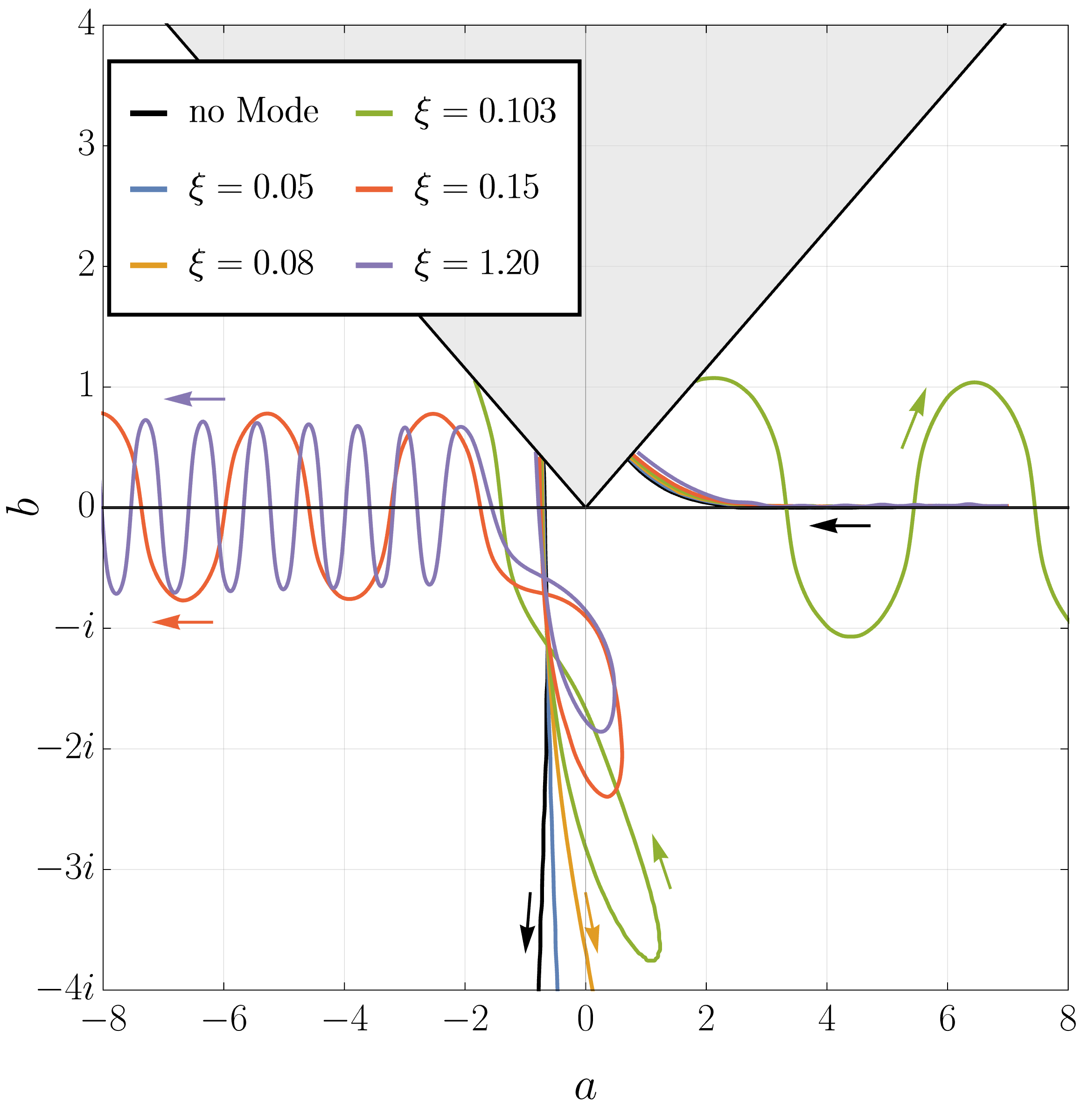}
    \caption{This figure shows the trajectories of the $2+1$ scattering with the $\eta_1$ mode excited. The black curve illustrates the BPS geodesic. The colored trajectories represent cases in which the 2-vortex was initially excited with its lowest mode. The blue line corresponds to the case used to measure  the spectral flow in the simulation. All the trajectories start from the positive $a$-direction ($b=0$, $a=7$).}
    \label{fig:eta1}
\end{figure}

We begin with the excitation of the lowest mode, $\eta_1$. This mode can be excited by the excitation of the lowest mode of the axially symmetric 2-vortex. In Figure~\ref{fig:eta1}, we show trajectories of the colliding vortices in the moduli space $a$-$b$-plane, for different values of the mode amplitude, $\xi$. The original BPS geodesic, where no mode is excited, is plotted as a black curve. 

We remind the reader that whenever the trajectory crosses the $a$-axis there is a $90^\circ$ scattering of two 1-vortices. Furthermore, for $a>0$ ($a<0$), two 1-vortices are on the l.h.s. (r.h.s).

Initially, the trajectories remain close to the BPS geodesic. In all cases, the trajectory explores the $b>0$ region and passes the $b=0$ line. This means that the 2-vortex splits horizontally and one of its 1-vortex constituents (the most right one) performs a right-angle scattering with the 1-vortex incoming from the r.h.s.. We can clearly see that afterwards the trajectories leave the BPS geodesic path even for small amplitudes of the mode. They bend toward growing $a$ and eventually enter the $a>0$ half-plane. Thus, the two 1-vortices off the $x$-axis and the 1-vortex on the $x$-axis exchange positions. The final state is a 1-vortex moving along the $x$-axis to the right, whereas two 1-vortices located off the $x$-axis move to the left. Therefore, we observe a {\it backscattering} of the 1-vortex.

For sufficiently high amplitudes, the trajectories remain in the $a>0$ region only for a finite time and cross the $a=0$ line for a second time. In the $a$-$b$-plot, we can observe a loop that is formed by the trajectory in the moduli space. Thus, the two 1-vortices off the $x$-axis and the 1-vortex on the axis exchange the positions again. 
Afterwards, there are two possible outcomes we have observed. In a small window around $\xi\sim 0.1$, the trajectories enter the $a>0$ half-plane again and pass the $b=0$ line many times while $a$ is increasing. This represents multi-bounce $90^\circ$ collisions between the two 1-vortices moving to the left and a single 1-vortex moving to the right (see the green curve in Figure~\ref{fig:eta1}). 
For higher amplitudes the trajectories remain in the $a<0$ half-plane and pass the $a$-axis multiple times as $a$ is decreasing. This corresponds to multi-bounce collisions between the two 1-vortices moving to the right and a single 1-vortex moving to the left.

The presented dynamics can be qualitatively explained as a mutual interplay between the mode-generated force and the Coriollis effect. While the Corriolis effect bends the trajectory, the attractive force generated by the $\eta_1$ mode, brings the 1-vortex constituents together. This leads to the formation of the observed trajectory loop. In the final state for high amplitude scattering, the multi-bounce structure between two 1-vortices is triggered by the excitation of the lowest mode of the axially symmetric 2-vortex. Of course, this mode is excited by the initial condition.

Notice that the shape of the trajectory can change under a change in the phase of the excitation, $i\omega t \mapsto i\omega t + i\varphi$. That this modifies the influence of the mode on the dynamics has already been noted in~\cite{Krusch:2024vuy}. For the cases presented in this paper, we observe a similar effect. However, the overall structure of the trajectories remains the same.

As a final check of the prediction for the spectral flow of the $\eta_1$ mode presented in section~\ref{sec:spectral-flow-along-the-geodesic}, we excited the $2$-vortex with a small amplitude ($\xi=0.05$, corresponding to the blue line in Figure~\ref{fig:eta1}). For this amplitude, the trajectory lies very close to the BPS geodesic, and thus we can reconstruct the spectral flow by measuring the frequency of the modes. For more details on how this is achieved in the simulation, we refer to the Appendix. In Figure~\ref{fig:flow}, the result is added as a light-gray line behind the blue line. We observe that the measurement from the simulation agrees very well with the prediction.

\subsection{$\eta_2$ mode}
\begin{figure}
    \centering
   \hspace*{-0.5cm} \includegraphics[width=1.03\linewidth]{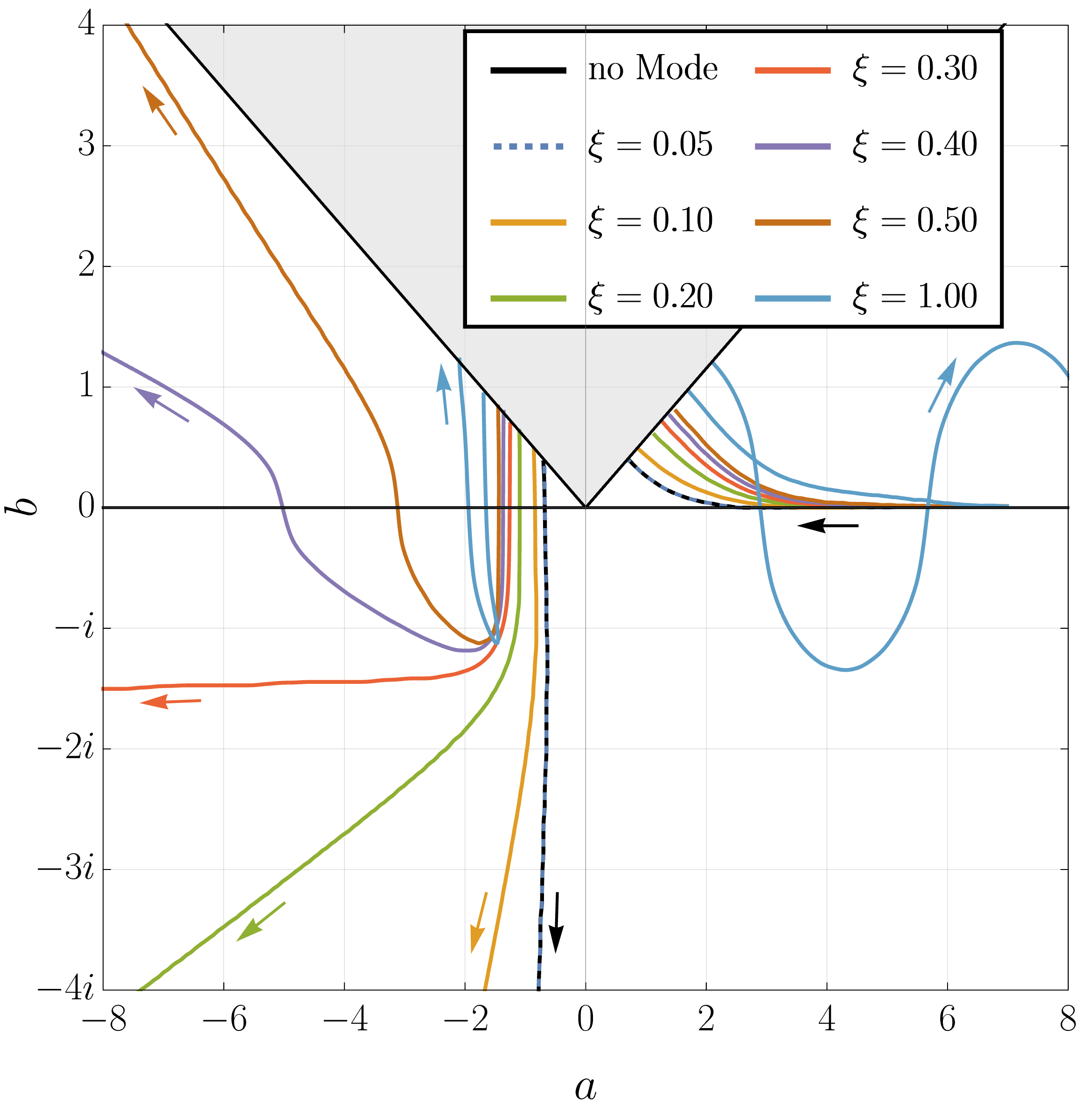}
    \caption{This figure shows the trajectories of the $2+1$ scattering with the $\eta_2$ mode excited. The black curve illustrates the BPS geodesic. The colored trajectories represent cases in which the 1-vortex is excited in the initial state. The blue dashed line corresponds to the case which we used to measure the spectral flow in the simulation. All the trajectories start from the positive $a$-direction ($b=0$, $a=7$).}
    \label{fig:eta2}
\end{figure} 

For very small amplitudes $\xi$, we see again only a tiny deformation of the BPS dynamics (see the blue dashed line in Figure~\ref{fig:eta2}, which agrees perfectly with the BPS geodesic in black). The trajectory in the $a$-$b$-plane crosses the $a$-axis only once. Thus, there is a single 
$90^\circ$ scattering between two of the 1-vortices. For increasing amplitudes, the trajectory is smoothly deformed and bends into the negative $a$ direction. This is again a consequence of the combined interplay between the repulsive mode-generated force and the Coriollis effect. The repulsive force reduces the $v_y$ velocity of the 1-vortices located off the $x$-axis. This is visible as a flattening of the trajectory as the amplitude increases (the red curve). Then, for a certain amplitude, $v_y=0$ and the two r.h.s. vortices asymptotically move parallel to the $x$-axis.

When the amplitude grows further, we enter the regime where the two vortices on the r.h.s. collide again. After this collision, all three 1-vortices are again located on the $x$-axis. 
Now, one of them moves to the left and two move to the right (the purple curve). So, we find the {\it passing through regime}.

If the amplitude of the mode is even higher, leading to a large enough repulsive force, the second 2-vortex collision occurs when the third vortex is still close. Then, after the 90$^\circ$ scattering of the right two vortices we can see a 90$^\circ$ scattering between the left two vortices. So, we observe an {\it interchange of the partners}. Now in the final state, the two vortices located on the l.h.s. perform a number of bounces. This is because of the excitation of the in-phase mode of the 2-vortex state. Thus, we have a bouncing 2-vortex state going to the left and a 1-vortex moving to the right. This is the {\it backscattering regime}.
This is represented in Figure~\ref{fig:eta2} by the light blue trajectories that crosses the $a$-axis (for $a>0$) several times. The actual number of bounces, and therefore the actual final state, depends chaotically on the initial data.

Again, as a check, we measure the spectral flow of the $\eta_2$ mode in the simulation. For this, we excite the $1$-vortex with a small amplitude ($\xi=0.05$, corresponding to the blue dashed line in Figure~\ref{fig:eta2}). We then track the frequency with respect to the $a$ and $b$ coordinates. The result is shown as a light-gray line behind the orange line in Figure~\ref{fig:flow}. We again see that the measurement from the simulation agrees very well with the prediction.

Summarized, it is worth to underline that although the dynamics between the excited 2- and 1-vortices reveals a chaotic pattern in the final state formations, see e.g., the chaotic bounces between two 1-vortices, from the point of the view of the $a$-$b$-plane, the trajectories behave in a rather well-organized manner. Definitely, the trajectories do not explore the moduli space in a random way. In particular, there are structures which seem to stabilize the trajectory. In the case of the $\eta_1$ mode, it is the circular orbit arising due to the attractive force, while in the case of the $\eta_2$ mode, it is the focal point triggered by the repulsive force.

\section{4-vortex scatterings}
A natural extension of this study is to include a fourth vortex. In this section, we present an analysis analogous to that of the 2+1 vortex scattering case for a few cases of 4-vortex configurations.

\subsection{Reduced moduli space}
The BPS 4-vortex configurations form an 8-dimensional moduli space of energetically equivalent solutions. As in the 3-vortex case, we restrict our analysis to solutions with the center of mass being located at the origin. Furthermore, we assume a subspace of configurations with additional symmetries, $x\to -x$ and $y\to -y$. This two-dimensional reduced moduli space, $\widetilde{\mathcal{M}}_4^{\rm CM}$, defined by the zeros of the scalar field is very non-trivial and for a full understanding it requires a neatly written parametrization.

One part of the moduli space can be parametrized by
\begin{align}
\label{4-v}
    &z_1=a+b\, ,\phantom{-} \hspace{1cm} z_2=a-b\, ,\nonumber\\ 
    &z_3=-a+b\, , \hspace{1cm}z_4=-a-b\, .
\end{align}
This choice describes vortices in the corners of a rectangle. Here, $a\in\mathbb{R}$ and $b\in i\mathbb{R}$ are the coordinates in the complex plane. Since the transformation $a\mapsto -a$ and $b\mapsto -b$ gives the same configuration (up to a relabeling), we use the restrictions $a \in \mathbb{R}_+$ and $b\in i\mathbb{R}_-$. For $b=-ia$ we obtain the square configurations. Furthermore, for $b=0$ ($a=0$), the configuration reduces to two charge-2 vortices located on the $x$-axis ($y$-axis). The solution with $a=b=0$ is the axially symmetric 4-vortex.

To get four vortices on the $x$-axis (or $y$-axis) we have to continuate the coordinates to $a\in \mathbb{R}_+$ and $b\in \mathbb{R}_+$ (and $a\in i\mathbb{R}_-$ and $b\in i\mathbb{R}_-$). These are collinear configurations. Note that there is an additional symmetry, $a\to b$, $b\to a$. Hence, the line $b=a$, defines a boundary of a region which has to be excluded to avoid double counting. This line describes a collinear configuration with the axially symmetric 2-vortex at the origin and two 1-vortices at equal distance $2a$.

\begin{figure}
   \includegraphics[width=1.02\linewidth]{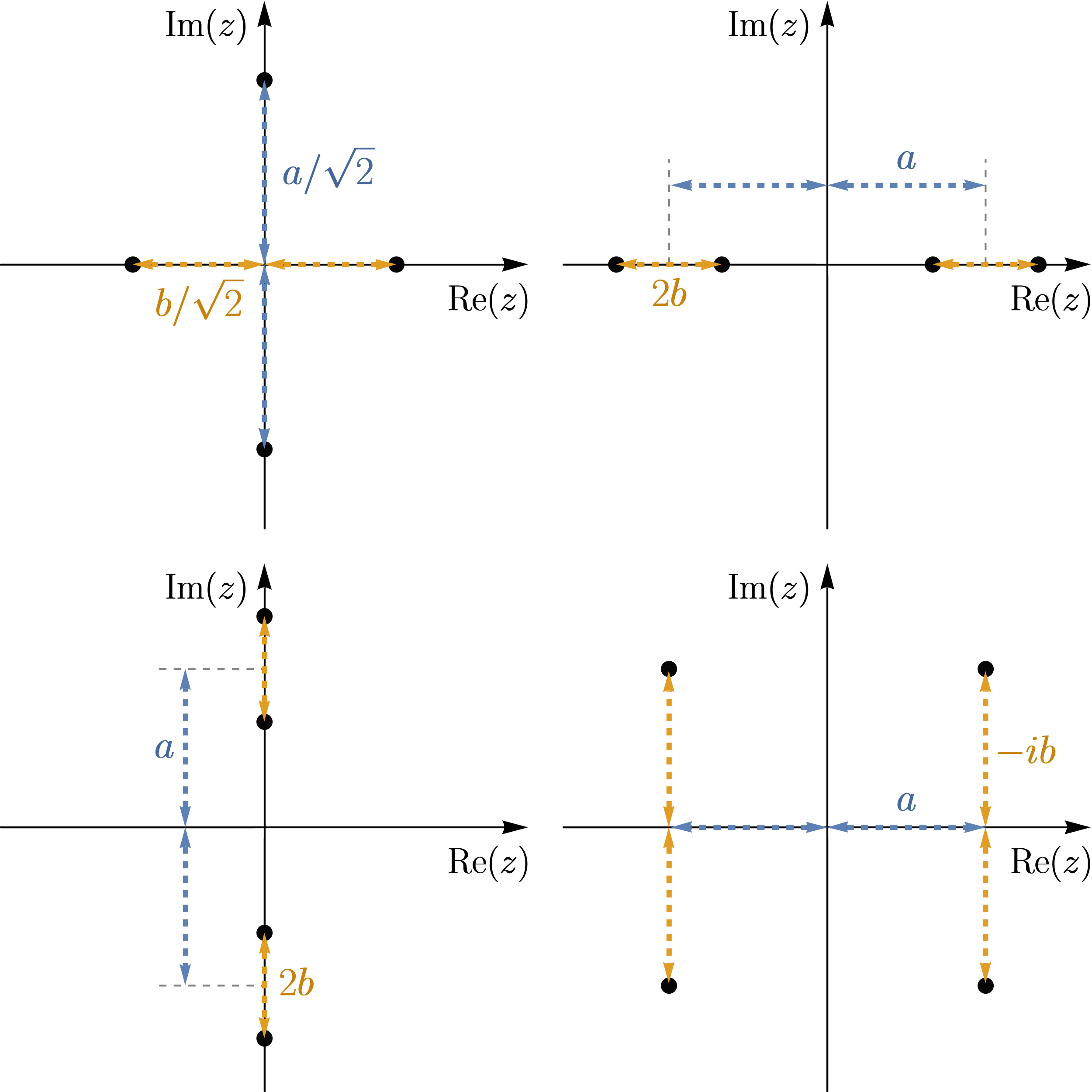}
    \caption{This figure illustrates the definition of the $a$ and $b$ values with respect to the distances between the vortices for the four different configurations. Top left: cross. Top right: collinear $x$-axis. Bottom left: collinear $y$-axis. Bottom right: rectangular.}
    \label{fig:n4_config}
\end{figure}

The second additional (and different) parameterization covers configurations that we called {\it cross configurations}, where two zeros are located on the $x$-axis and two on the $y$-axis 
\begin{align}
\label{eq:def-cross-parametrization}
    &z_1=\sqrt{2}\, a\, ,\phantom{-} \hspace{1cm} z_2=\sqrt{2}\, b\, ,\nonumber\\ 
    &z_3=-\sqrt{2}\, a\, , \hspace{1cm}z_4=-\sqrt{2}\, b\, ,
\end{align}
where $a\in i\mathbb{R}_-$ and $b\in \mathbb{R}_+$. These solutions are not captured by the parameterization~\eqref{4-v}. The factor $\sqrt{2}$ is chosen to guarantee that the vortices in the square and in the equal-arm cross configurations have the same distance from the origin. The parametrizations given in equations~\eqref{4-v} and~\eqref{eq:def-cross-parametrization} are illustrated in Figure~\ref{fig:n4_config}.\\

\begin{figure}
    \centering
    \includegraphics[width=1.03\linewidth]{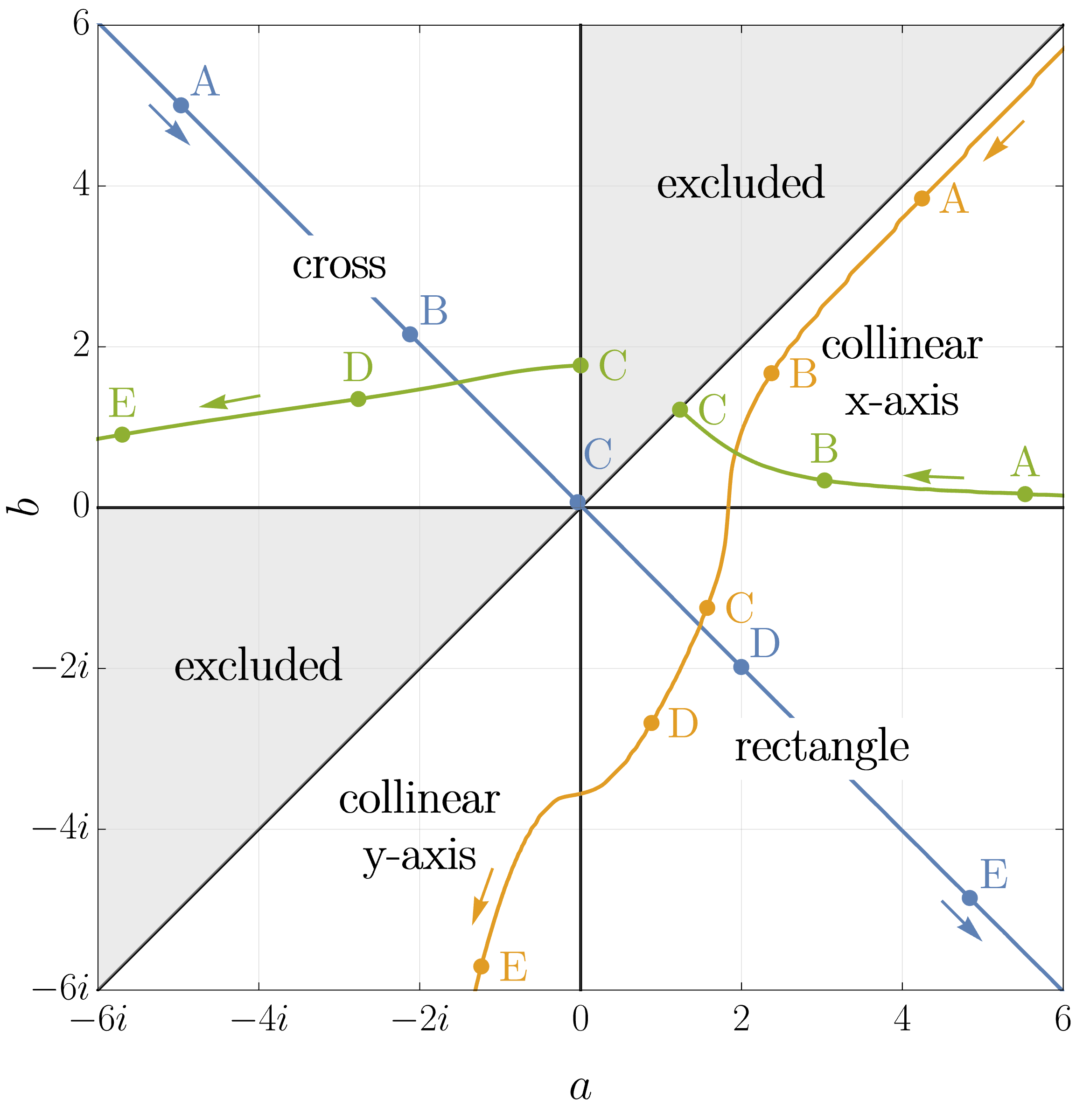}
    \caption{This figure shows the moduli space of 4-vortex configurations with the geodesics corresponding to the square (blue), collinear 1+2+1 (orange) and 2+2 vortex collisions (green). The labeled dots correspond to the time frames in the Figures~\ref{fig:energy_density_1+1+1+1},~\ref{fig:energy_density_1+2+1}, and~\ref{fig:energy_density_2+2}, respectively. The initial velocities of the moving vortices are $\abs{v}=0.01$ in all three cases.}
    \label{fig:moduli_space_4}
\end{figure}

\begin{figure*}
    \centering
    \includegraphics[width=\textwidth]{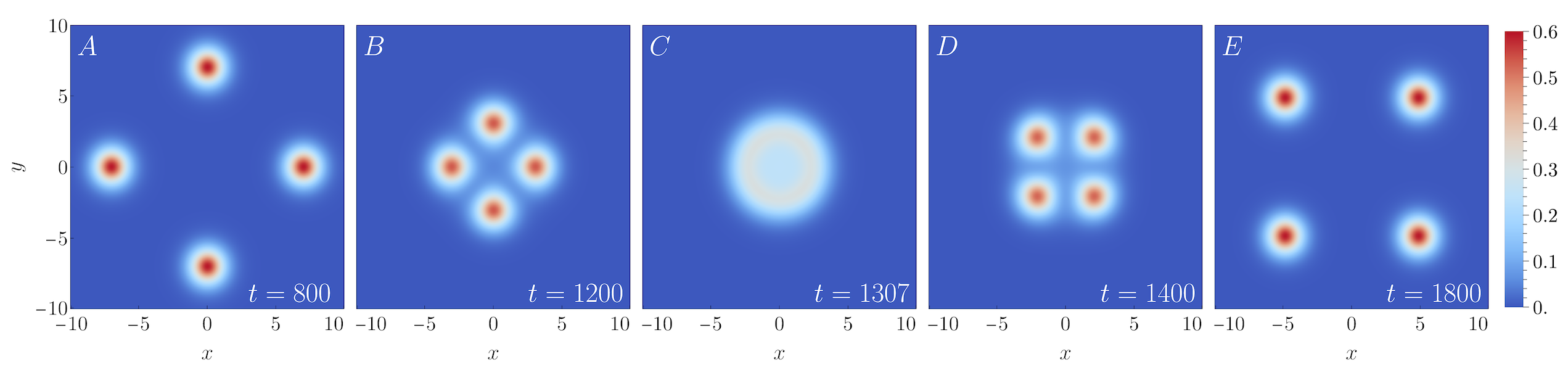}
    \caption{This figure displays five time frames of the energy density during the square scattering. The time frames correspond to the blue dots in Figure~\ref{fig:moduli_space_4}.}
    \label{fig:energy_density_1+1+1+1}

\vspace*{0.3cm}

    \centering
    \includegraphics[width=\textwidth]{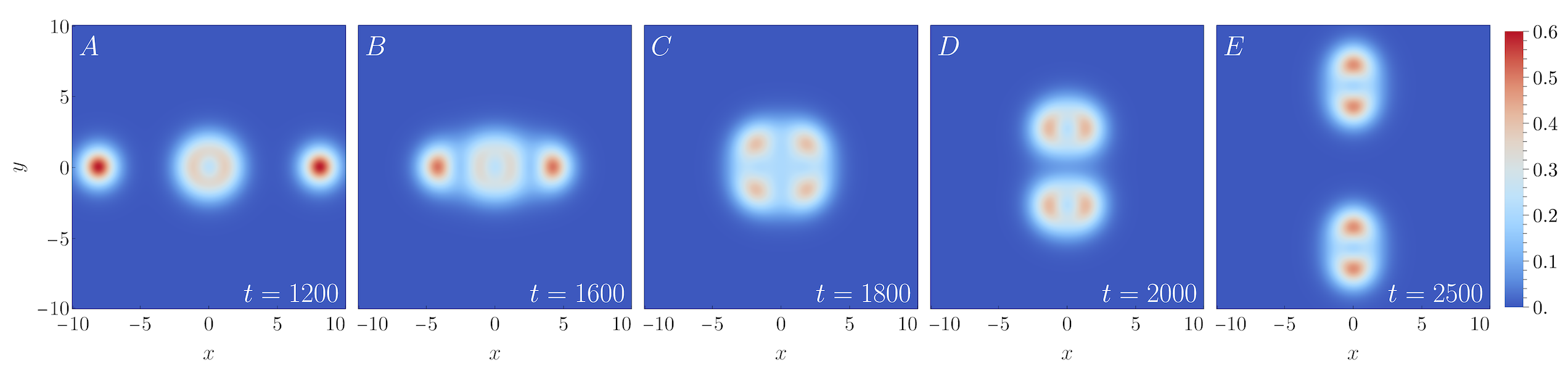}
    \caption{This figure displays five time frames of the energy density during the collinear 1+2+1 scattering. The time frames correspond to the orange dots in Figure~\ref{fig:moduli_space_4}.}
    \label{fig:energy_density_1+2+1}

\vspace*{0.3cm}

    \centering
    \includegraphics[width=\textwidth]{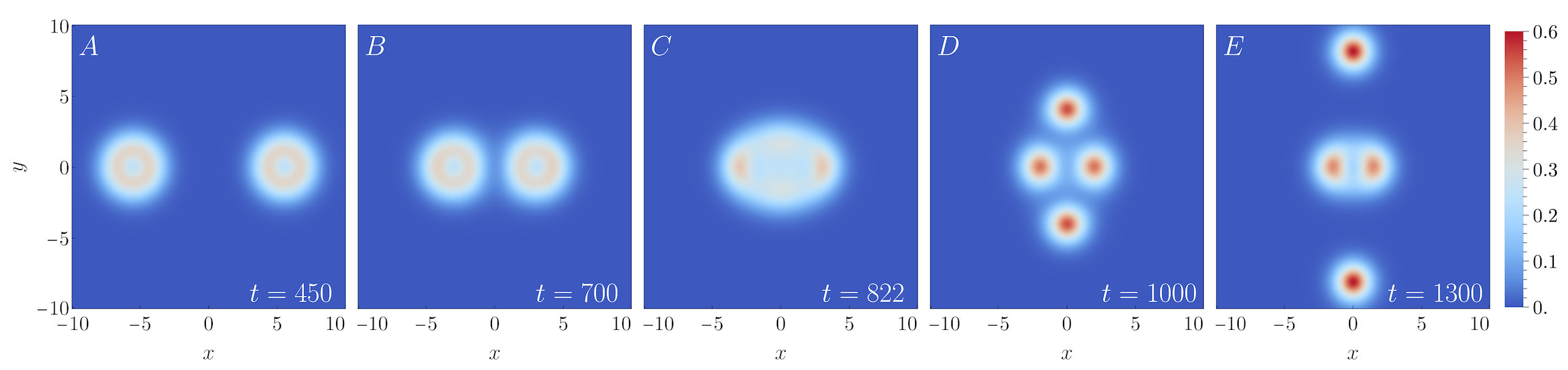}
    \caption{This figure displays five time frames of the energy density during the 2+2 scattering. The time frames correspond to the green dots in Figure~\ref{fig:moduli_space_4}.}
    \label{fig:energy_density_2+2}
\end{figure*}

In Figure~\ref{fig:moduli_space_4} we show the resulting moduli space. The shaded regions denote the part of the space that has to be excluded in order to avoid a double counting. In the figure, we are also showing three BPS geodesics corresponding to the square, collinear 1+2+1, and 2+2 scattering. We obtained these geodesics using numerical simulations with initial vortex velocities $\abs{v}=0.01$. In the Figures~\ref{fig:energy_density_1+1+1+1},~\ref{fig:energy_density_1+2+1}, and~\ref{fig:energy_density_2+2} we show five time frames of the corresponding energy density for each case. The animated versions of the energy density evolution can be found in the following video:\\
\url{https://youtu.be/o-0fQn-0wlE}

The blue line represents four 1-vortices located in the corners of a rotated square, $b=-ia$. As $a\to 0$, we pass the axial 4-vortex solution. The 1-vortices scatter under $45^\circ$~\cite{Arthur:1995eh} and form a square with the edges being parallel to the $x$- and $y$-axes. These configurations reveal an enhanced symmetry, which is the $90^\circ$ rotation. Therefore, we expect that even if we add excitations that don't break this symmetry, the four 1-vortices will evolve along the same geodesic.

The orange line denotes the evolution of the 1+2+1 collinear configuration, which initially has the axially symmetric 2-vortex at the origin and two outer 1-vortices symmetrically located on the $x$-axis. As $a$ decreases, the outer 1-vortices come closer to the origin and the central 2-vortex splits along the $x$-axis. This results in two $90^\circ$ scatterings between two pairs of two 1-vortices. The geodesic bends towards the region of the moduli space corresponding to rectangular configurations. Now, the 1-vortices in the pairs located in the upper and lower half-planes attract each other. This leads to two additional right-angle scatterings at $y=0$. Therefore, in the final configuration we find all 1-vortices located on $y$-axis. In the $a$-$b$-plane this is described by the geodesic entering the region with the collinear $y$-axis configurations.

The green line illustrates the head-on scattering between two charge-2 vortices. Initially, they are located on the $x$-axis ($a\to \infty$). As they come closer, they split along the $x$-axis, forming 4-vortex collinear configurations. At some point, the two inner 1-vortices collide in the origin, forming a single 2-vortex. This corresponds to hitting the boundary of the excluded region $b=a$ in the moduli space (point C in Figure~\ref{fig:moduli_space_4}). Subsequently, the configuration passes to the cross configurations with two (originally outer) 1-vortices on the $x$-axis and two (originally inner) 1-vortices on the $y$-axis. The vortices on the $x$-axis slowly approach each other and eventually meet. This results in a last $90^\circ$ scattering. Once again we find the collinear $y$-axis configurations in the final state.
 
We would like to underline that it is a numerically difficult task to find the geodesics for the 1+2+1 and 2+2 collisions. Since there are infinitely many energetically equivalent configurations and there is no additional symmetry, even a small change in the initial data can have an impact on the long-time evolution. Notice that our initial data is never perfect, that is, the incoming vortices have always a finite separation, which is not the true asymptotic solution for the ideal 1+2+1 or 2+2 configuration. Nevertheless, we simulated the scatterings for various initial separations and velocities and always got qualitatively very similar results. The trajectory of the 1+2+1 collision is especially stable. The geodesic corresponding to the 2+2 collision is for all initial separation and velocities qualitatively the same but changes its curvature in the cross region, resulting in an earlier (or later) passing to the y-collinear region. In any case, our accuracy is sufficient to study the main goal of the work which is the impact of the internal modes on the dynamics of the vortices.

\subsection{Spectral flow}
\begin{figure}
    \centering
    \includegraphics[width=1.0\linewidth]{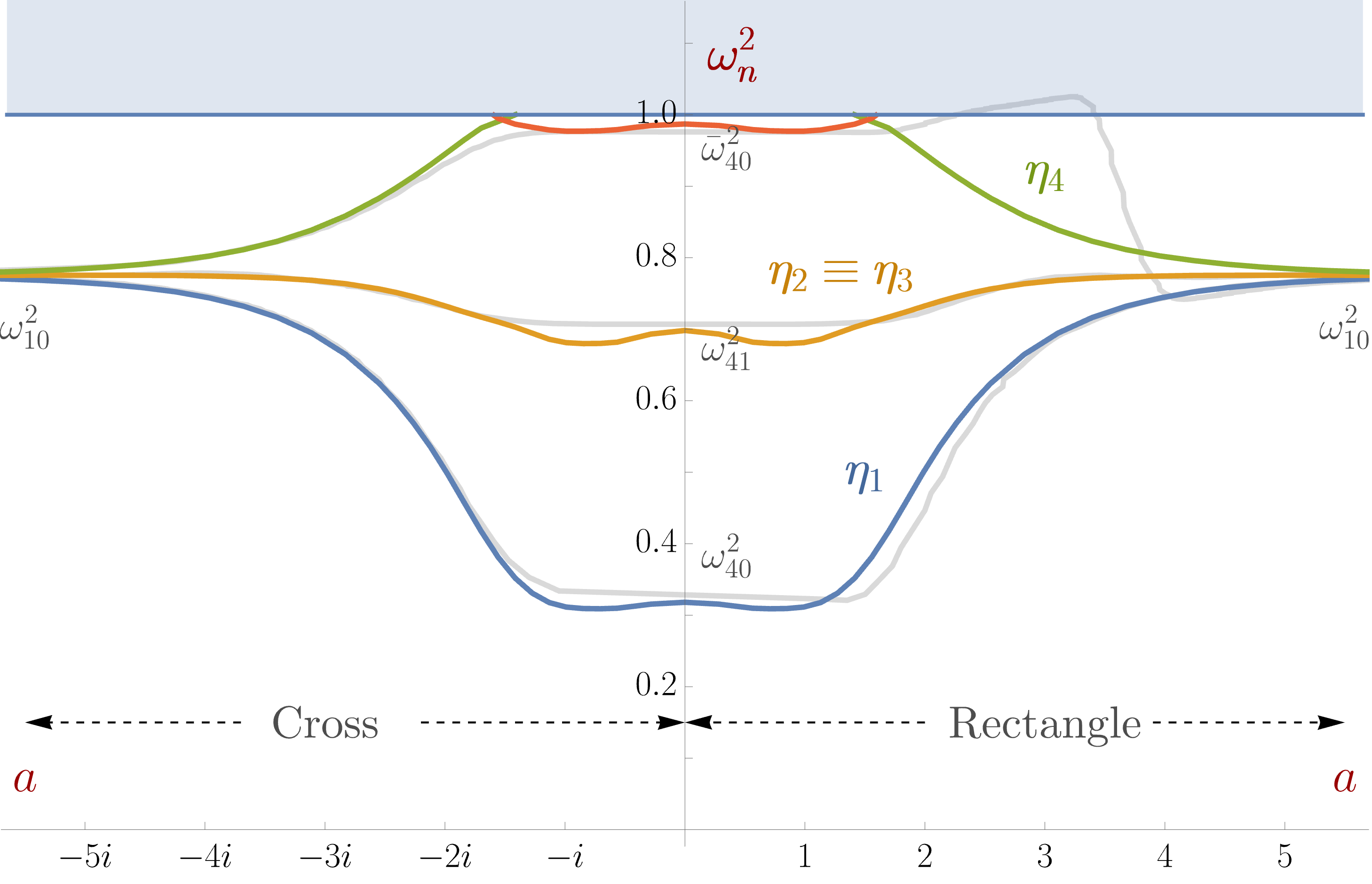}
    \caption{This figure shows the flow of the spectrum along the BPS geodesic path of the square collision. The light gray lines correspond to the measured frequencies obtained from the numerical simulations of the vortex collisions with the respective initial excitation.}
    \label{fig:flow_square}
\end{figure}
   
\begin{figure}
     \includegraphics[width=1.0\linewidth]{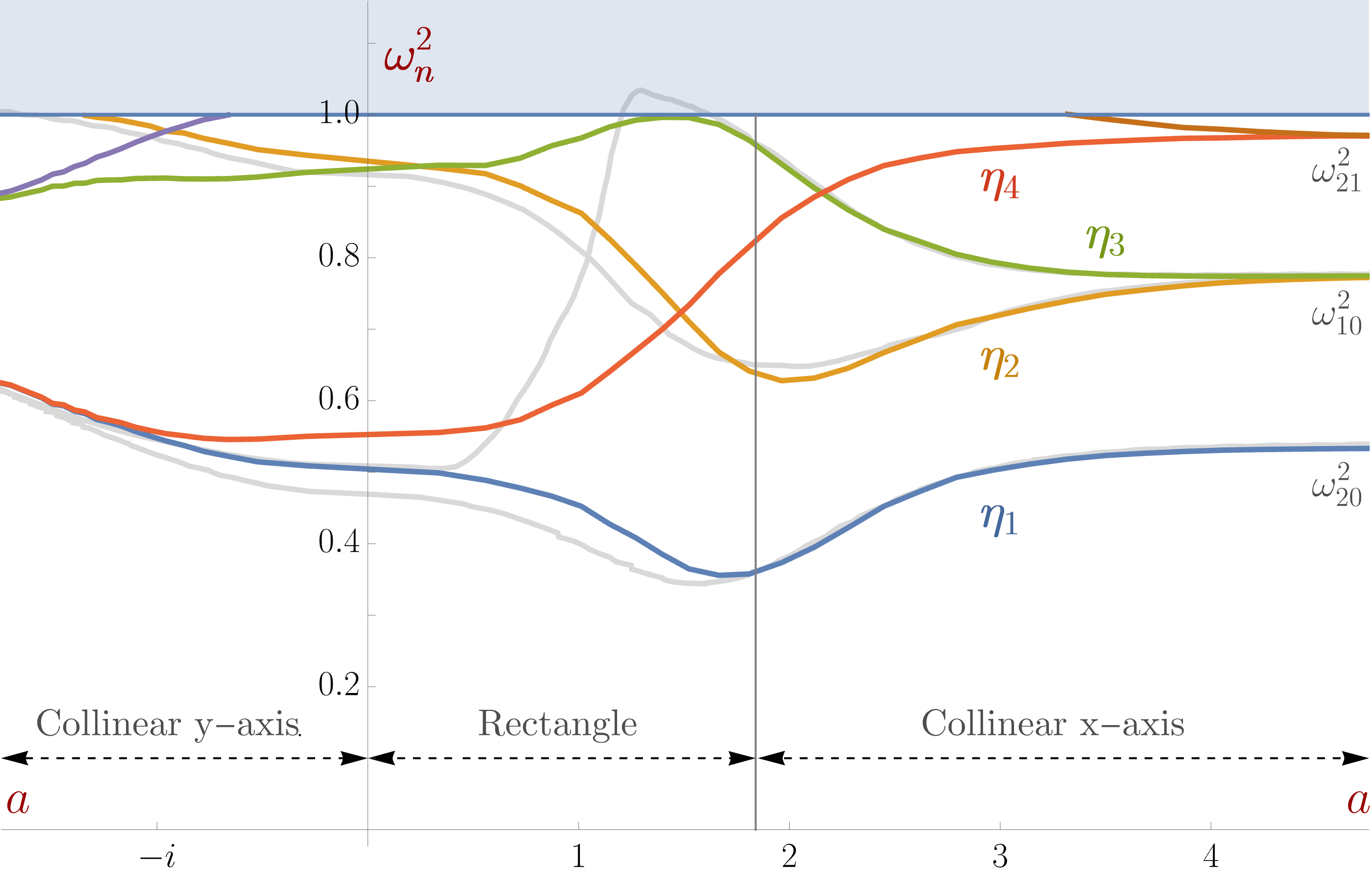}
    \caption{This figure shows the flow of the spectrum along the BPS geodesic path of the 1+2+1 collision. The light gray lines correspond to the measured frequencies obtained from the numerical simulations of the vortex collisions with the respective initial excitation.}
    \label{fig:flow_121}
\end{figure}

In Figures~\ref{fig:flow_square} and~\ref{fig:flow_121}, we present the flow of the spectral structure along the geodesics of the square and the head-on 1+2+1 collisions, respectively. 

From Figure~\ref{fig:flow_square}, we can see that the spectral flow for the most symmetric square collision reveals a rather simple and regular structure. Initially, there are four far separated 1-vortices. This amounts to four different (linearly independent) superpositions of the 1-vortex shape modes. The symmetric superposition, 
\begin{equation}
    \xi_1=\frac{1}{2} \left(\ket{1}+\ket{2}+\ket{3} +\ket{4} \right)\, ,
\end{equation}
leads to the lowest frequency mode $\eta_1$ (blue curve). Its frequency gets smaller as the 1-vortices approach each other. Two other orthogonal superpositions 
\begin{align}
    \xi_2=&\frac{1}{2} \left(\ket{1}+\ket{2}-\ket{3} -\ket{4}\right)\, , \\
    \xi_3=&\frac{1}{2} \left(\ket{1}-\ket{2}-\ket{3} +\ket{4}\right)\, ,
\end{align}
excite the two degenerated modes $\eta_{2,3}$, for which the frequencies change only slightly along the geodesic (orange). The last orthogonal superposition
\begin{equation}
    \label{eq:xi4-superposition}
    \xi_4=\frac{1}{2} \left(\ket{1}-\ket{2}+\ket{3} -\ket{4}\right)\, ,
\end{equation}
combines to the highest frequency mode $\eta_4$ (green). Its frequency increases as the vortices come closer. It seems that at $a\approx 1.4142$ in the rectangle regime (or similarly at $a\approx -1.4142i$ in the cross case) it hits the mass threshold. Hence, contrary to the first three modes, this mode probably does not evolve to one of the modes of the axially symmetric 4-vortex. There is another, the fifth mode, that emerges from the highest mode of the axial 4-vortex (red) and it disappears in the continuum at $a\approx 1.59$ (or similarly at $a\approx -1.59i$). However, we must underline that within our numerical precision we cannot exclude the possibility that these two modes form one mode that exists for all $a$. 

Due to symmetry reasons, the flow of the spectrum after the $45^\circ$ scattering is a mirror reflection.

As seen in Figure~\ref{fig:flow_121}, the change of the spectral structure along the 1+2+1 geodesic revels a much more complex behavior. Initially, there are two shape modes of two 1-vortices and three modes of the central axial 2-vortex. The lowest mode of the 2-vortex excites the lowest frequency mode $\eta_1$ (blue curve). Again, its frequency decreases as the vortices are coming closer to each other. The minimal value is reached at the point when the configuration passes from the $x$-collinear to rectangular region, that is, when the two left and the two right 1-vortices are on top of each other. Then, the frequency grows approaching the mode with a symmetric superposition of four 1-vortex shape modes. The higher degenerate modes of the 2-vortex excites the $\eta_4$ and the $\eta_5$ mode (red and brown curves, respectively). The frequency of the $\eta_4$ mode monotonically decreases along the geodesic until $a$ becomes complex negative, which corresponds to the $y$-collinear part. There, the mode smoothly joins with a superposition of four 1-vortex shape modes. The $\eta_5$ mode quickly crosses the mass threshold (brown curve).

The antisymmetric and symmetric superpositions of the shape mode of the two outer 1-vortices excite the $\eta_{2}$ and $\eta_3$ mode respectively (orange and green). Their flow along the geodesic is very non-trivial with some mode crossings. The $\eta_3$ mode approaches the mass threshold very closely and, at least within our accuracy, never disappears in the continuum. By measuring the frequency in the numerical simulation of the collision of the very weakly excited vortices (for which we see only tiny modification of the BPS trajectory), we can find that the mode is close enough to the mass threshold to disappear (see light gray line behind the green line). Afterwards it continues with the lowest mode (blue).
The $\eta_2$ mode hits the mass threshold in the collinear $y$-axis region. This eventually acts as a spectral wall.

\subsection{Dynamics}
As in the 2+1 vortex scatterings, we use numerical simulations to study the influence of the respective modes on the vortex trajectories. The numerical implementation follows the same procedure as in the 2+1 case. In all cases, the initial velocity of the Lorentz-boosted vortices is $\abs{v}=0.01$.

\interfootnotelinepenalty=10000
First of all, we must notice that the two-dimensional restricted moduli space $\widetilde{\mathcal{M}}_4^{\rm CM}$ contains configurations with a rather large symmetry. This should be respected by any initial data. Otherwise, the dynamics leaves $\widetilde{\mathcal{M}}_4^{\rm CM}$ and explores other directions in the full moduli space $\mathcal{M}_4^{\rm CM}$.\footnote{In Figures~\ref{fig:flow_square} and~\ref{fig:flow_121}, some gray lines were indeed obtained from excitations that break the symmetry. However, we kept the amplitudes of these excitations rather small, so that the resulting deformed path remains close to the BPS geodesic trajectory and does not significantly deviate from the two dimensional subspace, i.e., the $a$-$b$-plane.} In other words, some of the modes do not have an $x\to -x$ symmetry. In order to study the impact of these modes on the BPS geodesics, and especially plot it as a trajectory in a configuration space, one should consider a subspace respecting only $y\to -y$ symmetry, which is a complicated three-dimensional manifold. This goes beyond the scope of the present work. We will restrict our analysis to cases that don't leave the two-dimensional moduli space $\widetilde{\mathcal{M}}_4^{\rm CM}$.
\interfootnotelinepenalty=0

We begin with the square scattering. Here, the lowest mode, $\eta_1$, and the highest mode, $\eta_4$, respect the symmetries of $\widetilde{\mathcal{M}}_4^{\rm CM}$. In addition, excitation of the $\eta_1$ mode preserves the $45^\circ$ rotation symmetry. Thus, the dynamics must still go along the BPS geodesic. As in all previously known cases, the excitation of the lowest frequency mode generates an attractive force, which prefers the vortices to be on top of each other. Indeed, the frequency decreases as the 1-vortices approach each other. This results in bouncing solutions, where the 1-vortices go back and forth {\it along the BPS-trajectory} and pass the axially symmetric 4-vortex many times. The actual number of bounces and the resulting final scattering angle by $45^\circ$ chaotically depend on the initial data (velocity, amplitude of the mode, and the phase). 

\begin{figure}
    \centering
   \hspace*{-0.5cm} \includegraphics[width=1.03\linewidth]{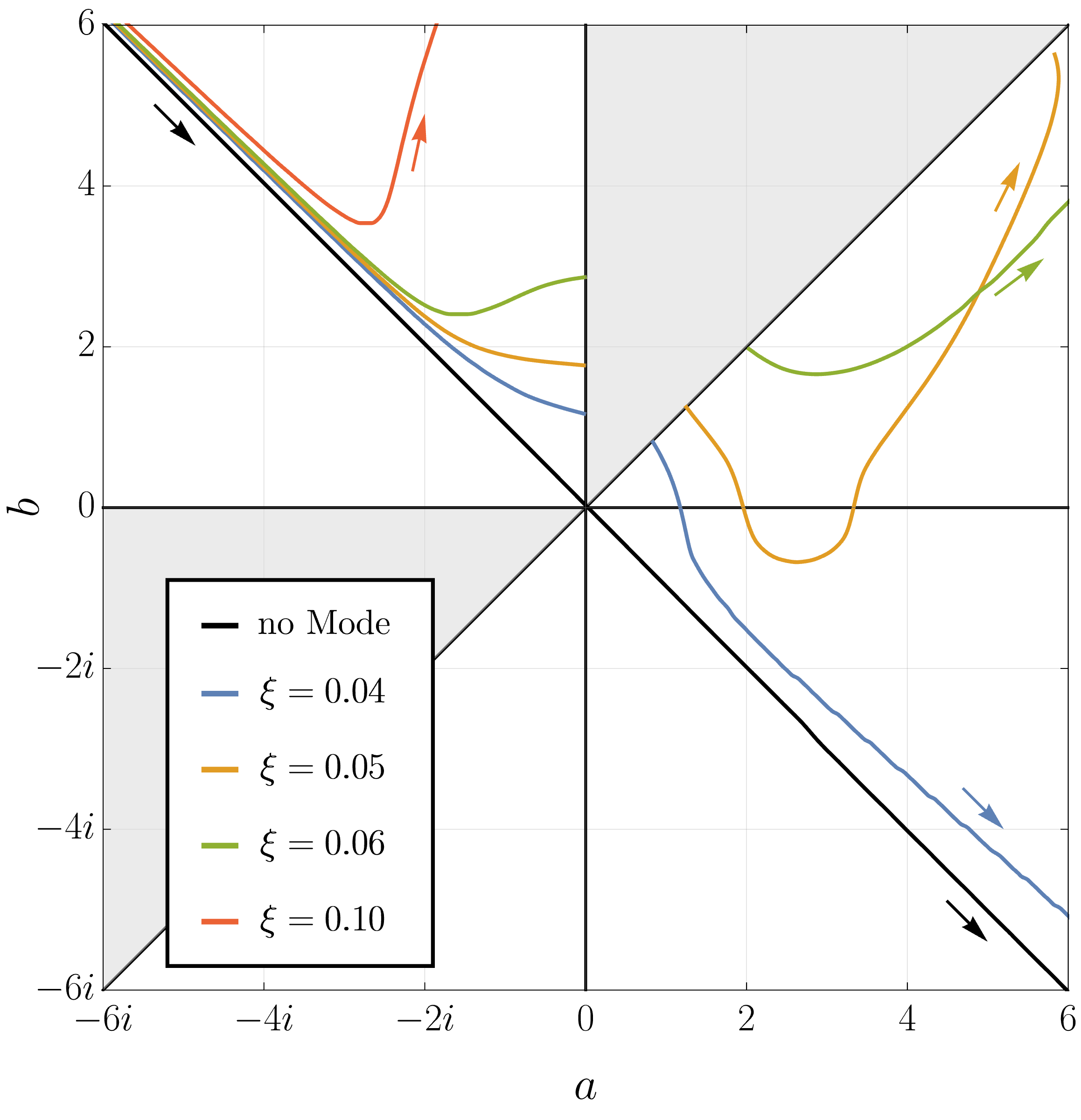}
    \caption{This figure shows the trajectories of the 4-vortex square scattering with the $\eta_4$ mode excited. The black curve illustrates the BPS geodesic. The colored trajectories correspond to cases in which all four 1-vortices are excited using the $\xi_4$ superposition given in equation~\eqref{eq:xi4-superposition}. The blue line corresponds to the case which we used to measure the spectral flow in the simulation.}
    \label{fig:square-eta4}
\end{figure}

The first main difference, resulting from the excitation of the $\eta_4$ mode, is the breaking of the $45^\circ$ rotation symmetry of the initial configuration. This means that the trajectory leaves the BPS geodesic during the time evolution even for small excitation amplitudes $\xi$. The second main difference is that such a mode generates a repulsive force. Both effects are visible in Figure~\ref{fig:square-eta4}, where we plot the actual trajectories in the restricted moduli space $\mathcal{M}_4^{\rm CM}$ for several values of the initial amplitude $\xi$. For small amplitudes we observe only a small deviation from the BPS geodesic where the final state has almost the square symmetry. Note that the two 1-vortices located on the $y$-axis collide (trajectory passes $a=0$), leading to the formation of collinear states along the $x$-axis. For moderate amplitudes, the inner two vortices bounce, which can be seen by the trajectory passing the $a$ axis twice (orange line). As the amplitude further grows, this bounce disappears (green line). Eventually, for sufficiently large amplitudes, we see strong repulsion (red line). The trajectory never passes to the collinear region. However, as we remarked, the repelled vortices do not remain on the BPS geodesic.

\begin{figure}
    \centering
   \hspace*{-0.5cm} \includegraphics[width=1.03\linewidth]{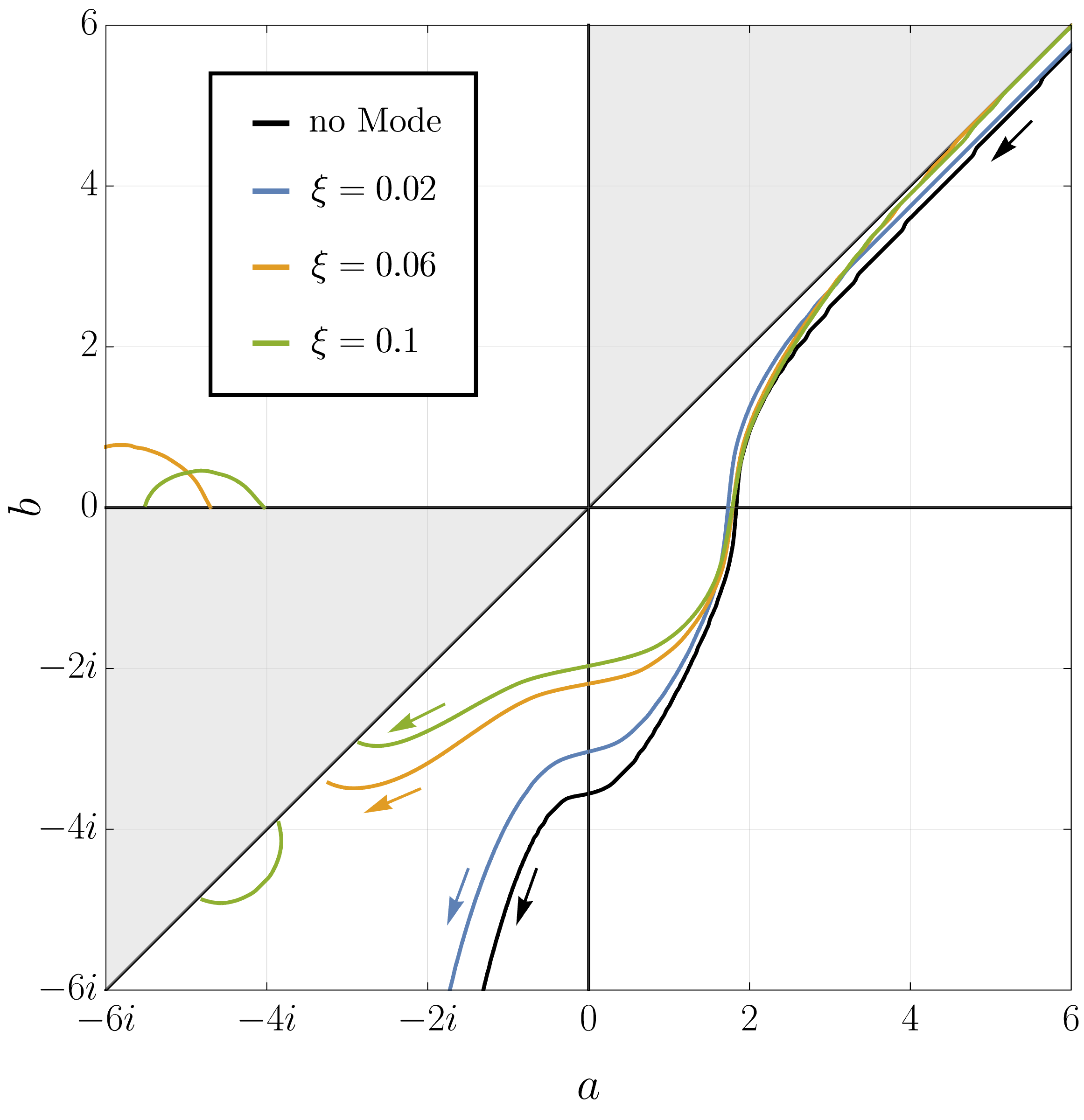}
    \caption{This figure shows the trajectories of the collinear 1+2+1 scattering with the $\eta_1$ mode excited. The black curve illustrates the BPS geodesic. The colored trajectories correspond to cases in which only the 2-vortex in the origin is excited. The blue line corresponds to the case which we used to measure the spectral flow in the simulation.}
 \label{fig:1+2+1-eta1}
    \vspace*{0.3cm}
    
     \hspace*{-0.5cm} \includegraphics[width=1.03\linewidth]{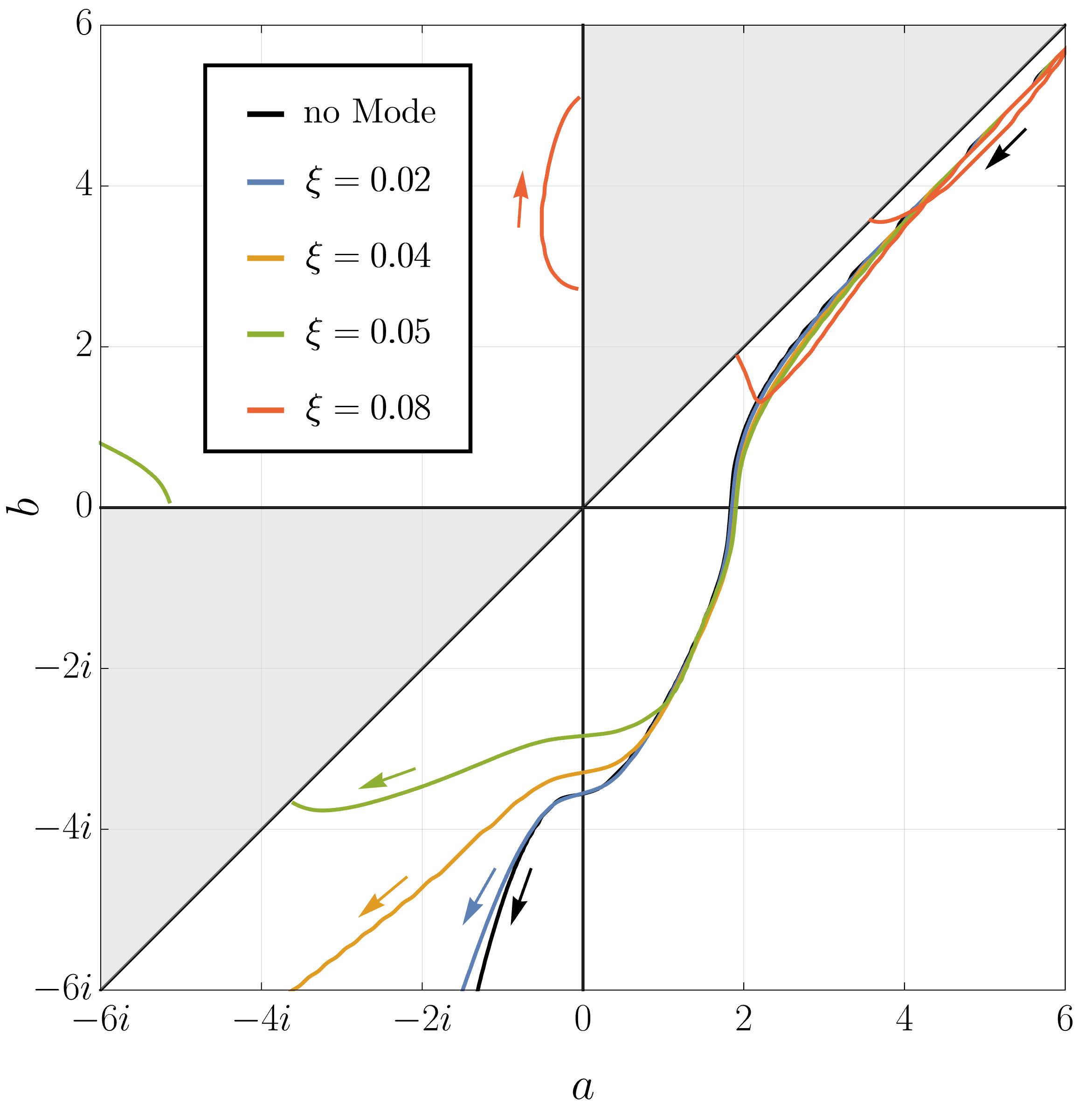}
    \caption{This figure shows the trajectories of the collinear 1+2+1 scattering with the $\eta_3$ mode excited. The black curve illustrates the BPS geodesic. The colored trajectories correspond to cases in which only the two 1-vortices are excited. The blue line corresponds to the case which we used to measure the spectral flow in the simulation.}
    \label{fig:1+2+1-eta3}
\end{figure} 

In the 1+2+1 collisions, the excitation of two modes respects the symmetries of the configurations forming the restricted moduli space. They are $\eta_1$ and $\eta_3$, which are triggered by the excitation of the lowest shape mode of the axial 2-vortex and the in-phase superposition of the modes of two 1-vortices, respectively.

As always, $\eta_1$ leads to the appearance of an attractive force, which together with the Coriollis effect, modifies the BPS geodesic (see Figure~\ref{fig:1+2+1-eta1}). As the amplitude increases, the trajectory goes back and forth from the $y$-collinear regime to the cross configurations. This reflects multiple head-on scatterings of two 1-vortices (at the origin), while the other 1-vortices are scattered to infinity along the $y$-axis.

The excitation of the $\eta_3$ mode generates a repulsive force, see for example the red trajectory in Figure~\ref{fig:1+2+1-eta3}. This is because, although the frequency initially slightly decreases as we initially go along the BPS geodesic, it strongly rises in the rectangular regime. For moderate amplitudes, the trajectory again passes from the $y$-collinear to the cross regime. This means that, as before, one pair of 1-vortices goes to infinities along the $x$-axis, while the second pair does it along the $y$-axis. In the numerical simulation we observed that the two vortices located around the origin in the final state can scatter again. This can be explained by the frequency measured in the simulation, indicated by the light gray line behind the green $\eta_3$ line in Figure~\ref{fig:flow_121}. In the simulation, the $\eta_3$ mode reaches the mass threshold and disappears. Afterwards, a small amplitude of the lowest $\eta_1$ mode remains. This excitation may arise from a non-linear interaction of the massive modes and also from the energy transfer from the zero mode (kinetic motion) to the massive mode. The lowest mode then induces an attraction between the two inner vortices, leading to the observed right-angle scattering around the origin.

\begin{figure}
    \centering
   \hspace*{-0.5cm} \includegraphics[width=1.03\linewidth]{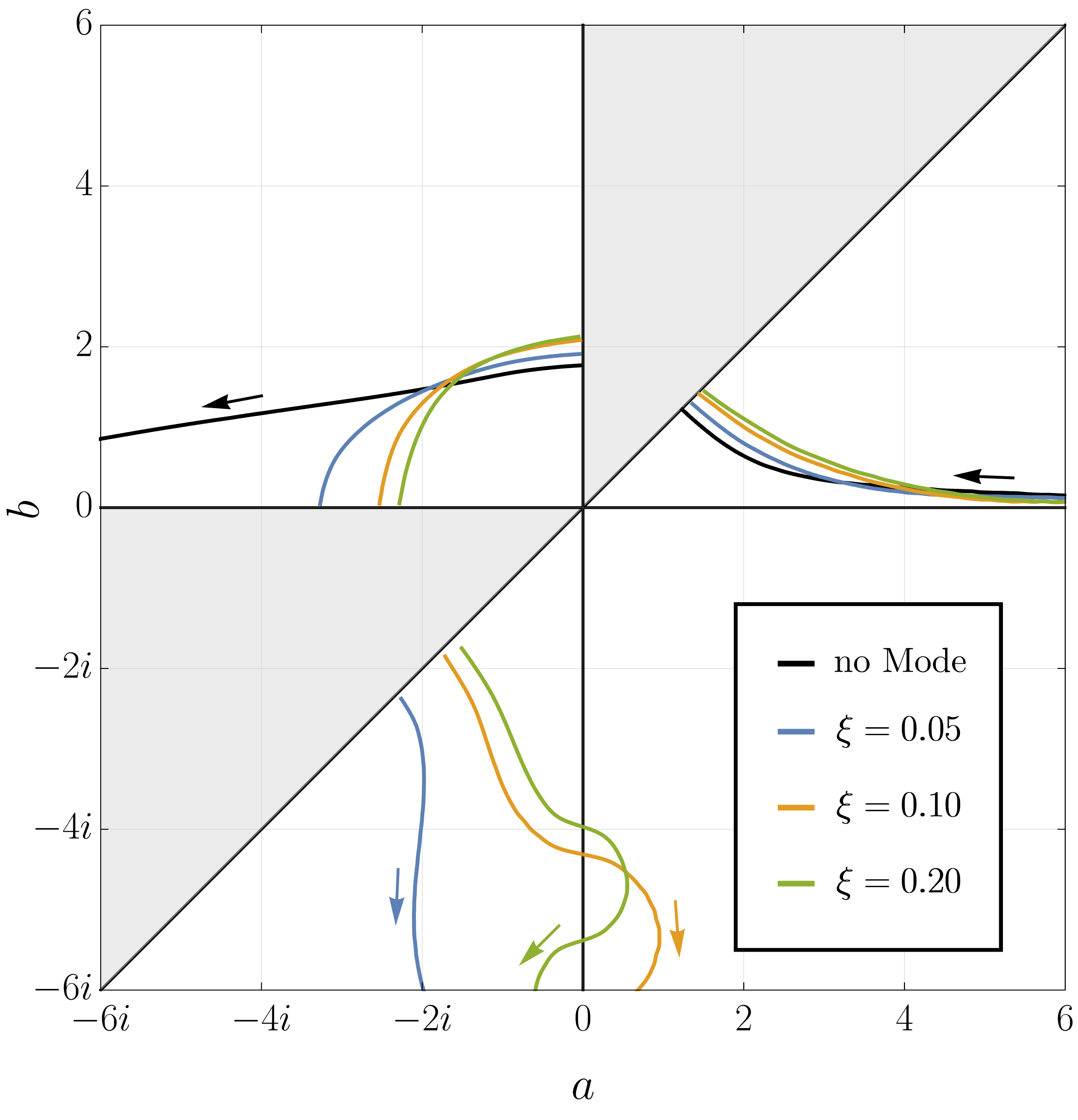}
    \caption{This figure shows the trajectories of the 2+2 scattering with both 2-vortices excited with its lowest mode. The black curve illustrates the BPS geodesic.}
    \label{fig:2+2}
\end{figure}

For the 2+2 scattering, there is one mode that preserves the symmetry of the reduced moduli space, which can be excited by exciting the lowest 2-vortex modes with the same phase. From Figure~\ref{fig:2+2} we observe that this mode is attractive and bends the trajectory in the $a<0$ half-plane toward the collinear $y$-axis region. The final state consists of two vortices moving in the positive $y$-direction and two vortices moving in the negative $y$-direction.
For sufficiently large amplitudes, the final configuration are two bouncing vortex pairs moving apart along the $y$-axis. For very high amplitudes, the bouncing amplitude is rather small, and the full process resembles two 2-vortices scattering by $90^\circ$.

\section{Conclusion and Outlook}

In this work, we investigated how the dynamics of vortices in the Abelian-Higgs model at critical coupling in the winding number $n=3$ and $n=4$ sectors is affected by the excitation of the internal modes. We focused in particular on cases without enhanced symmetry. In the first step, we numerically determined the BPS geodesics for different scattering scenarios of three and four vortices. Then, we considered scatterings with the modes excited. 

It turns out that the dynamics changes profoundly when the vortices are initially excited. In fact, it reveals a much more sophisticated behavior than the standard geodesic flow on the moduli space of energetically equivalent solutions. This marks a significant difference compared to the previously studied cases with enhanced symmetry~\cite{Krusch:2024vuy, AlonsoIzquierdo:2024nbn, Alonso-Izquierdo:2024fpw,Alonso-Izquierdo:2025suz}.
In general, there are two main sources of modifications of the BPS dynamics:  the well understood {\it mode-generated force}, which emerges as a potential on the moduli space, and the less understood {\it Coriolis effect}, i.e. the modification of the moduli space metric induced by the amplitude of the mode. Both effects have a substantial impact on multi-vortex collisions.

In the scatterings with enhanced symmetry, also respected by the excited mode, the system evolves along the original BPS geodesic path. The reason is that the symmetry of the initial data prohibits any change of the geodesic. This is the case for head-on collisions in the $n=2$ sector, the equilateral triangle and collinear scatterings in the $n=3$ sector and the square scatterings in the $n=4$ sector. However, because of the appearance of the mode-generated force, the actual dynamics does change. It can go back and forth along the BPS geodesic and is fully governed by the flow of the mode along the original BPS geodesic. Depending on the behavior of the frequency of the mode, an attractive or repulsive force can emerge. This may lead to {\it chaotic} multi-bounce solutions, as, e.g., the excitation of the lowest mode in the $n=2$ head-on collisions and in the scattering of four 1-vortices along the square geometry.

In a generic case, we observed that the original BPS trajectory gets modified if a mode is excited. It makes the evolution even more chaotic. Nevertheless, the modified geodesic does not randomly cover the full moduli space but explores, in a quite regular form, only some part of it. This is again governed by the mode-generated force and by the Coriollis effect. 

A general tendency is a break-up of multi-vortex configurations into a collection of two-vortex pairs.
For exited vortices, these may be quite stable long-lived bound states where the one-vortices bounce many times.\\

The huge impact of the excited internal modes on the dynamics of the BPS vortices (and also cosmic strings) may have important phenomenological consequences.

There are many sources that can excite the massive internal modes of the vortices. For example during a phase transition they can already emerge with excitations or thermal fluctuations can excite the modes. Furthermore, interactions with radiation, in collisions with other (anti)vortices, or via coupling to an oscillating background~\cite{Kitajima:2025nml} are natural sources for the dynamical excitation of the massive vortex modes (see also~\cite{Blanco-Pillado:2021jad} for a discussion of the dynamical excitation of the global vortex modes).
In general, solitonic bound modes decay very slowly~\cite{Manton_1997}, which is also known to be true for vortices~\cite{Alonso-Izquierdo:2024tjc} and for global (axionic) cosmic strings~\cite{Kibble:1976sj, Vilenkin:1982ks}, although parametric resonances can lead to a faster transition of energy between the radial shape modes and the transverse zero modes~\cite{Blanco-Pillado:2022axf}. Hence, in a generic situation, the vortices or cosmic strings should be considered in an excited state.
Importantly, the dynamics of axionic strings as well as the spectrum of the Goldstone mode radiation produced by the strings is essential for realistic estimations of the abundance of axions~\cite{Weinberg:1977ma, Wilczek:1977pj, Preskill:1982cy} and its contribution to the today's dark matter.  

The dynamics of axionic strings has been extensively studied within the thin-wall approximation~\cite{Vilenkin:1986ku, Garfinkle:1988yi, Battye:1993jv, Battye:1995hw, Drew:2019mzc}. Furthermore, axionic strings can also be connected by axionic domain walls~\cite{Vilenkin:1982ks,Kibble:1982dd}, which modifies their dynamics.\footnote{Notice that recently in~\cite{Dvali:2025xur} it was shown that the fermionic zero modes, which determine whether or not domain walls are attached to strings, are set by the QCD chiral condensate.} However, within all investigations on the dynamics, it is remarkable that massive normal modes are usually not taken into account.
As suggested in~\cite{Hindmarsh:2021mnl}, the existence of excited massive modes can also be the main reason for a disagreement between the effective Nambu-Goto theory~\cite{Blanco-Pillado:2013qja} (where no bound modes are included) and fully field theoretic (lattice) simulations~\cite{Hindmarsh:2017qff} of local string loops. Our findings qualitatively confirm that the geodesic dynamics of 3- and 4-vortices, which can be viewed as a lower-dimensional version of the Nambu-Goto dynamics of strings, can be drastically affected by excitation of the massive mode.\\

In order to get a better qualitative or even quantitative description of the collisions of BPS-vortices, e.g., given by a collective coordinate model, the flow of the spectral structure over the full (restricted) moduli space has to be computed. Furthermore, a better understanding of the Coriollis effect is required. So far, its impact on the vortex dynamics is only known in the single vortex sector~\cite{Miguelez-Caballero:2025xfq}.

Undoubtedly, the importance of the role of the modes in the dynamics of other higher-dimensional solitons should be investigated further. We expect that the dynamics of non-BPS vortices~\cite{Stuart:1994yc, Speight:1996px, Speight:2025qrr}, non-Abelian vortices~\cite{Eto:2011pj},~chiral skyrmions \cite{Theodorou:2025seu}, (cosmic) strings, and BPS monopoles~\cite{Manton:1981mp, Manton:1988ba, Bachmaier:2025jaz}, will also be very sensitive to the excitation of internal modes. 
Furthermore, it was recently noticed~\cite{Dvali:2025sog} that the excitation of ``memory modes" may have a significant impact on the dynamics of black holes. Therefore, studying mode-generated changes in the dynamics of solitons may also provide new insights into black hole dynamics.

\section*{Acknowledgement}

The authors thank D. Chiego, G. Dvali, and N. Manton for discussion and remarks.

A. A. I.  and A. W. acknowledge support from the Spanish Ministerio de Ciencia e Innovacion (MCIN) with funding from the grant PID2023-148409NB-I00 MTM.

\section*{Appendix}
\subsection*{\normalsize Appendix A: $\widetilde{\mathcal{M}}^{\rm CM}_4$ moduli space}
The moduli space of energetically equivalent BPS solutions in the charge-4 sector is defined by the positions of the charge-1 components, i.e., the zeros of the scalar field. They can be expressed as the roots of a fourth order complex polynomial
\begin{align}
        P_4(z)&=(z-z_1)(z-z_2)(z-z_3)(z-z_4)\nonumber \\
        &=z^4+w_3z^3+w_2z^2+w_1z+w_0\, ,
\end{align}
where
\begin{align}
    w_3&=-(z_1+z_2+z_3+z_4)\, , \\
    w_2&=z_1z_2+z_1z_3+z_1z_4+z_2z_3+z_2z_4+z_3z_4\, , \\
    w_1&=-(z_2z_3z_4+z_1z_3z_4+z_1z_2z_4+z_1z_2z_3)\, , \\
    w_0&=z_1z_2z_3z_4\, ,
\end{align}
are complex numbers. As in the 3-vortex case, the motion of the center of mass decouples from the other degrees of freedom. Thus, without loss of generality, we can locate it at the origin, $w_3=0$. In the next step, we can assume that $w_1=0$ and the two other coefficients $w_0$ and $w_2$ take real values. This brings us to a two-dimensional subspace $\widetilde{\mathcal{M}}_4^{\rm CM}$ generated by the polynomials
\begin{equation}
    \tilde{P}_4^{\rm{CM}}= z^4+w_2z^2+w_0\,,
\end{equation}
where $w_0,w_2 \in \mathbb{R}$. One can verify that this corresponds to the vortex solutions with $x\to -x$ and $y \to -y$ symmetry.

For example, for the configurations parameterized by~\eqref{4-v}, which covers the rectangular and $x$- or $y$-collinear solutions, we find 
\begin{equation}
    w_2=-2(a^2+b^2)\,, \;\;\; w_0=(a^2-b^2)^2\,,
\end{equation}
with $w_0 \in \mathbb{R}_+$ and $w_2 \in \mathbb{R}$. For the cross solutions, we obtain 
\begin{equation}
    w_2=-2(a^2+b^2)\,, \;\;\; w_0=4a^2b^2\,,
\end{equation}
where $w_0 \in \mathbb{R}_-$ and $w_2 \in \mathbb{R}$. 

\subsection*{\normalsize Appendix B: Numerical methods}
\textbf{Initial Configuration.} For all multi-vortex configurations considered in this work, we employed a product ansatz for the scalar field. For instance, a static configuration of a 2-vortex and a 1-vortex is given by
\begin{align}
    \phi(x,y)=\phi^{(1)}(x-x_1,y-y_1)\,\phi^{(2)}(x-x_2,y-y_2)\, ,
\end{align}
where $\phi^{(n)}$ denotes the vortex solution given in equation~\eqref{eq:vortex-ansatz}. We fixed the center of mass to lie at the origin, which implies $x_1=-2x_2$. We further chose $y_1=y_2=0$, with $x_1=14$ and $x_2=-7$.
Since the theory is Higgsed, the gauge field is massive and therefore suppressed outside the vortex core. Consequently, it is a good approximation to use a summation ansatz for the gauge field,
\begin{align}
    A_\mu(x,y)=A^{(1)}_\mu(x-x_1,y-y_1)+A^{(2)}_\mu(x-x_2,y-y_2)\, .
\end{align}

To endow the vortices with initial velocities (along the $x$-axis), we simply Lorentz boost the configuration by replacing $x-x_n$ with $\gamma_n(x-v_n t-x_n)$, where $v_n$ denotes the velocity and $\gamma_n$ the corresponding Lorentz factor. For the gauge field, it is necessary to additionally Lorentz transform the vector field itself by $A^{(n)}_t\mapsto -v_n\gamma_n A^{(n)}_x$ and $A^{(n)}_x\mapsto \gamma_n A^{(n)}_x$.

For the cases with total winding $4$, we proceeded analogously. The initial distance between the vortices was always chosen to be at least $20$. Specifically, for the 1+1+1+1, 1+2+1, and 2+2 configurations, the vortices were initially placed at $\lbrace(15,0);(0,15);(-15,0);(0,-15)\rbrace$, $\lbrace(-20,0);(0,0);(20,0)\rbrace$, and $\lbrace(-10,0);(10,0)\rbrace$, respectively.\\

\textbf{Numerical time integration.}
The time integration was performed using the second-order Runge-Kutta method (RK2) (more details on this method can be found in~\cite{Figueroa:2020rrl}).

Due to the gauge freedom in the Abelian-Higgs model, it is necessary to fix the gauge by choosing an appropriate gauge condition. Since we are Lorentz-boosting the vortices, the Lorenz gauge, $\partial_\mu A^\mu = 0$, is a practical choice, because the initial configuration satisfies it. 

Following~\cite{Krusch:2024vuy}, we implemented natural boundary conditions, combined with the adiabatic damping method of~\cite{Gleiser:1999tj}. Specifically, outside of a radius of $25$, a friction term was added to the field equations, with a coefficient that grows as a Gaussian function of the radial distance. 

The simulations were carried out on a two-dimensional square lattice with $x, y \in [-30, 30]$ and a spatial spacing of $0.05$. The temporal step size was chosen as $0.025$, and the system was evolved up to a final time of $t = 4000$ (in cases for which the vortices hit the boundary, we decreased the investigated time range).

For some $4$-vortex simulations, we used a bigger lattice of size $x, y \in [-40, 40]$, due to the large initial separation. For these cases the radius for the friction term was increased to $35$.

For all simulations presented in this paper, we used the lattice specifications and initial velocities described above. In addition, we performed several numerical checks by varying the lattice size, lattice resolution, vortex separation distance, and initial velocities. Changing the lattice size or resolution produced qualitatively similar results, with only minor quantitative differences in the overall dynamics.

Increasing the initial velocities led to slight modifications of the trajectories, particularly when the velocity reached values of order $\sim 0.1$. At these higher velocities, we observed that internal modes can become excited during the collision, even when no modes are initially excited.

For smaller initial separation distances, the simulations became numerically unstable. The initial overlap between the vortices led to instabilities in the late-time evolution ($t\gtrsim 1000$).
In contrast, we obtained the most stable and reliable results for larger separations, $d \gtrsim 20$. This observation motivates our choice of the minimal initial separation distance mentioned above.
We also note that for very large initial separation distances, the vortices remain effectively non-interacting for a long period of time. In the case of the charge-$2$ vortex, it may occur that the single winding-$2$ zero splits into two winding-$1$ zeros. This splitting can be triggered by small numerical fluctuations. Although we observed this effect in some simulations, the resulting separation distance between the two zeros remains small and does not affect our qualitative analysis.\\

\textbf{Vortex position measurement.}
To determine the positions of the vortex centers, we first compute the winding number for each grid cell. To this end, we associate an angle with each value of the complex scalar field $\phi$. We then sum the angle differences around the four lattice points forming a given grid cell. If this sum is approximately $2\pi n$, the cell contains an $n$-vortex. Numerically, for a configuration with total charge $n$, we identify the vortex centers by selecting the $n$ grid cells with the largest winding numbers. In addition, we must exclude lines where the angle discontinuously jumps between $0$ and $2\pi$.\\

\textbf{Spectral flow measurement of BPS configurations.}
The spectral flows shown in Figures~\ref{fig:flow},~\ref{fig:flow_square}, and~\ref{fig:flow_121} were obtained by tracking the trajectories of the unexcited vortex centers in the simulations, selecting approximately $50$ frames that are considered as representatives of the overall evolution. Using these frames, static BPS solutions were constructed by fixing the vortex centers to the positions from the selected frames. This construction was performed by applying a gradient flow method on a $2500\times 2500$ grid of the region $(x,y)\in [-25,25]$. The relaxation was performed until the difference between the energies of the relaxed configurations and the exact BPS solutions was smaller than 0.01, starting from an initial configuration that was generated by using a generalized Abrikosov ansatz. Full details of this procedure can be found in section 3 of~\cite{Alonso-Izquierdo:2023cua}. Subsequently, the spectrum was computed for each static configuration by discretizing the vortex small-fluctuation operator and diagonalizing a submatrix extracted from the full matrix associated with the initial grid. The size of the submatrix was increased until the eigenvalues stabilized, typically for submatrix orders of $100$, $125$, or $150$. All numerical calculations were implemented in C++, using the Eigen library for matrix diagonalization tasks.\\

\textbf{Frequency measurement.}
The spectral flows of the modes shown in Figures~\ref{fig:flow},~\ref{fig:flow_square}, and~\ref{fig:flow_121} were computed using the method described in the previous paragraph. As an additional consistency check, we independently extracted the mode frequencies directly from the time evolution in the numerical simulations.
Since the shape modes correspond to excitations that periodically transfer energy between different components of the total energy, this feature can be used to determine the frequency. In particular, it is sufficient to monitor a single energy component to measure the frequency of the energy exchange. 

We decided to analyze the total potential energy of the system. Notice that in certain cases, when modes were (or became) excited out of phase, it was necessary to compute the potential energy only within one half-plane of the lattice. From the time dependence of the potential energy, we determined the oscillation period by measuring the time interval between neighboring peaks. The corresponding frequency can be computed from this period time. To reduce numerical noise, we applied a Gaussian filter to the data, resulting in a smoothed frequency profile. The measured spectral flow curves are shown as light gray lines in Figures~\ref{fig:flow},~\ref{fig:flow_square}, and~\ref{fig:flow_121}.

Notice that the excitation of the modes in the simulation leads to deviations of the vortex trajectories from the BPS geodesic. Therefore, we do not expect the measured spectral flow to coincide perfectly with the predicted behavior. In particular, for the $\eta_{2,3}$ modes in the square collision and the $\eta_2$ mode in the 1+2+1 collision, the trajectories even leave the two-dimensional moduli space $\widetilde{\mathcal{M}}_4^{\rm CM}$.
Nevertheless, for the small excitation amplitudes considered here, this effect is rather small and the measured frequencies are in very good agreement with the results obtained using the above described method of the spectral flow measurement for BPS configurations.\\

\textbf{Programming language for the simulations.}
The numerical simulations were performed using Python, together with the NumPy library and the JIT compiler Numba~\cite{Numba}. Numba significantly improves the computational performance, by translating the Python code into optimized machine code and enabling efficient parallelization. All figures were generated using Mathematica.

\bibliography{ref}

\end{document}